\PassOptionsToPackage{table}{xcolor}

\documentclass[3p]{elsarticle}

\usepackage{hyperref}
\usepackage{todonotes}

\usepackage{amsfonts}
\usepackage{subfigure}
\usepackage{amsmath}
\usepackage{acronym}
\usepackage{multirow}
\usepackage{array}
\usepackage{hanging}
\usepackage{xcolor}
\usepackage{xspace}
\usepackage{floatrow}
\usepackage[export]{adjustbox}
\usepackage{tikz}
\usepackage{epstopdf}

\usetikzlibrary{shapes,arrows,positioning}
\usetikzlibrary{arrows.meta}

\acrodef{HH}{household}
\acrodef{HC}{household cluster}
\acrodef{SLA}{statistical local area}
\acrodef{DZN}{destination zone}
\acrodef{CD}{census collection district}
\acrodef{WG}{working group}
\acrodef{ABS}{Australian Bureau of Statistics}
\acrodef{BITRE}{Australian Bureau of Infrastructure, Transport and Regional Economics}

\newcommand{\new}[1] {{\color[rgb]{0.,0.,0.} #1}}

\newcommand{\acemod}{\textsc{Ace}Mod\xspace}

\newcolumntype{L}[1]{>{\raggedright\let\newline\\\arraybackslash\hspace{0pt}}m{#1}}
\newcolumntype{C}[1]{>{\centering\let\newline\\\arraybackslash\hspace{0pt}}m{#1}}
\newcolumntype{R}[1]{>{\raggedleft\let\newline\\\arraybackslash\hspace{0pt}}m{#1}}

\journal{Simulation Modelling Practice and Theory}

\def\arraystretch{1.3}
\begin{document}

\begin{frontmatter}

\title{Investigating Spatiotemporal Dynamics and Synchrony of Influenza Epidemics in Australia: An Agent-Based Modelling Approach}

\author{Oliver M. Cliff$^{1}$, Nathan Harding$^{1}$, Mahendra Piraveenan$^{1}$, E. Yagmur Erten$^{1,2}$, Manoj Gambhir$^{3}$, and Mikhail Prokopenko$^{1,4}$}
\address{$^{1}$ \quad Centre for Complex Systems, Faculty of Engineering and IT, University of Sydney, Sydney, NSW 2006, Australia\\
$^{2}$ \quad Department of Evolutionary Biology and Environmental Studies, University of Zurich, Winterthurerstrasse 190, 8057 Zurich, Switzerland\\
$^{3}$ \quad IBM Research, Melbourne, Australia\\
$^{4}$ \quad Marie Bashir Institute for Infectious Diseases and Biosecurity, University of Sydney, Westmead, NSW 2145, Australia
}

\begin{abstract}
{In this paper we present \acemod, an agent-based modelling framework for studying influenza epidemics in Australia. The simulator is designed to analyse the spatiotemporal spread of contagion and influenza spatial synchrony across the nation. The individual-based epidemiological model accounts for mobility (worker and student commuting) patterns and human interactions derived from the 2006 Australian census and other national data sources. The high-precision simulation comprises 19.8 million stochastically generated software agents and traces the dynamics of influenza viral infection and transmission at several scales. Using this approach, we are able to synthesise epidemics in Australia with varying outbreak locations and severity. For each scenario, we investigate the spatiotemporal profiles of these epidemics, both qualitatively and quantitatively, via incidence curves, prevalence choropleths, and epidemic synchrony. This analysis exemplifies the nature of influenza pandemics within Australia and facilitates future planning of effective intervention, mitigation and crisis management strategies.}
\end{abstract}

\begin{keyword}
\texttt{computational epidemiology\sep agent-based modelling\sep discrete-time simulation\sep influenza\sep epidemics\sep synchrony\sep demographics
}
\end{keyword}

\end{frontmatter}

\section{Introduction}


{Early recognition and curtailing of outbreaks of infectious diseases is crucial to policy making and public health. Without efficient intervention, international travel can result in the spread of highly infectious pathogens around the globe within weeks of the initiation of the outbreak, as evidenced by the 2009 H1N1 (swine flu) pandemic. Moreover, infectious diseases are large burdens on the economy. For instance, a 2007 study estimated that the annual cost of seasonal influenza to the health system in Australia was AU\$828-884 million~\cite{newall2007economic}. Effective strategies for mitigating outbreaks should include a suitable combination of immunisation, vaccination, and palliative care~\cite{lanzieri2014modeling,lopman2012understanding,simmons2013duration} and planning these strategies requires reliable forecasting through simulation of various "what if?" scenarios.} Without efficient intervention, highly infectious pathogens, such as swine flu, result in international travel carrying the virus around the globe within weeks to months of the initiation of the outbreak, causing a worldwide public health emergency~\cite{Longini2005,Ferguson,nsoesie2012sensitivity,nsoesie2014systematic}.

{Intergration of large-scale datasets and explicit simulations of entire populations down to the scale of single individuals has considerably improved the accuracy of epidemiological models~\cite{pastor}.} Furthermore, an effective intervention requires not only an accurate modelling and surveillance of the disease dynamics, but also {implementation} 
of the underlying contact networks and their topologies. This requires a detailed theoretical and practical understanding of the interplay between epidemic processes, mobility patterns (especially over long distances), and population data. By running multiple computer simulations while varying the sources of infection, 
{we can then ascertain} the average social and health impact, as well as ``zoom in'' on specific pathways and patterns of epidemics.

While there are currently many useful models and tools to assist in the analysis and prediction of contagion processes, none are adequate in terms of precise and integrated tracking, predicting and mitigating epidemics at the national-level for Australia. It remains a challenge to generate and simulate a realistic and dynamic contact network, accounting for both mobility and human interactions representing the demographics of Australia through an individual-based epidemiological model. Moreover, as evidenced by Cauchemez et al.~\cite{cauchemez2011role}, modeling the transmission of respiratory diseases must specifically account for structuring of schools, grades, and classes due to the ``back-and-forth waves of transmission between the school, the community, and the household.'' These challenges are particularly important for investigating epidemics in Australia, where the population is concentrated mainly around urban areas.

%
We have addressed this need by developing \acemod, the \emph{Australian Census-based Epidemic Model} that employs a discrete-time and stochastic agent-based model to investigate complex outbreak scenarios at various spatiotemporal levels. The simulator comprises 19.8 million software agents where each agent contains a set of attributes of an anonymous individual. The agents are generated such that the distributions at multiple scales concur with key demographic statistics of the 2006 Australian census data. The next layer of the model includes mobility patterns (with respect to work, study and other activities) in order to characterise potential interactions between spatially distributed agents. The final layer describes local transmission dynamics by combining agent health characteristics, such as susceptibility and immunity to diseases, with a natural history model for influenza. The simulation runs in 12 hour cycles (``day'' and ``night'') over the course of an epidemic such that agents interact in different social mixing groups depending on the cycle. Given a surrogate population generated in this way, we run multiple instances of each scenario, varying the disease infectiousness and {outbreak locations} to give insight into the spread of influenza pandemics around Australia.

\new{The novel aspects of AceMod include (i) the spatial fidelity of the stochastically generated population, calibrated to Australian Census and the Australian Bureau of Statistics (ABS) datasets that utilise multi-scale distributions over school data and hierarchical mixing groups; (ii) its refined models for the transmission and natural history of the simulated infection, based on the latest available epidemiological studies; and (iii) the focus on measures of spatiotemporal complexity of influenza epidemics.}

We evaluate the spread of influenza in space and time through Australia via incidence curves, prevalence choropleths, and epidemic synchrony. Incidence curves characterise the disease by plotting the number of newly \emph{ill} individuals (i.e., the incidence) at each time step. {Here, ill refers to infected agents that are showing symptoms.} This illustrates the severity of the disease over time and can be used
{to compare different pathogens or mitigation strategies.}
Then, in order to qualitatively study the evolving spatial distribution of the epidemic over time, we record {the percentage of ill individuals (i.e., the prevalence) in each community} at a given time step and give snapshots of this distribution through choropleths taken at key times of the epidemic. Finally, we examine spatial hierarchies in disease prevalence by analysing the variance {in the timing of}
the epidemic peaks in each community (i.e., the synchrony). {The spatiotemporal synchrony of disease spread between communities correlates to the size of the communities, suggesting the disease transmission is more associated with social connectivity rather than geographic distance.}

\section{Related work} \label{sec:background}

Well-established stochastic models of epidemics often use SIR and SEIR differential equations for the population dynamics of susceptible (S), exposed (E), infectious (I), and removed/recovered (R) individuals~\cite{keeling2011modeling}. 
{While these models are suitable for analysing the general behaviour of an epidemic on larger scales, with a focus on global variables, they do not allow us to make accurate predictions at a finer resolution.}

In many scenarios it is beneficial to trace the dynamics in a {more fine-grained} way especially during the initial or final stages of an outbreak, when person-to-person transmission processes dominate~\cite{GermannKadauEtAl2006}. As a result, stochastic agent-based discrete-time simulation models were developed to capture how the uncertainty in disease diffusion varies in different social groups and affects the overall analysis and predictions of epidemics~\cite{GermannKadauEtAl2006,Longini2005}. These models have been used to assess vaccination and antiviral prophylaxis strategies on a local level, highlighting importance of detailed modelling of contact patterns~\cite{Halloran2002,Longini2004,GermannKadauEtAl2006}. These models further allow us to investigate strategies at large-scale regional and national levels for containing an emerging pandemic influenza strain at its source~\cite{Longini2005,Ferguson}.

{Early efforts include an agent-based simulation by Elveback et al., where they modeled a small artificial population to represent a community and defined multiple contexts in which individuals interact, in order to explore the disease spread.}
Later works~\cite{Longini2005,GermannKadauEtAl2006} leveraged increased computational power to further this methodology, considering populations on a larger scale based on exact census data. \new{A suite of scalable agent-based modelling software simulators was described in a series of papers\cite{eubank2004modelling, barrett2008episimdemics, bisset2009epifast, barrett2010integrated, bisset2012simulating}. Simdemics was designed as a general purpose simulation environment supporting 300 million agents, aimed at modelling pandemic planning and response to a range of crisis scenarios, such as H1N1/H5N1 pandemics and bio-terrorism events \cite{barrett2010integrated}. A number of simulators were built upon Simdemics framework. For example, a high-performance computing modelling environment, EpiSims, designed as a distributed discrete event simulator, modelled disease transmission via sub-groups of people in a given location. The location objects were distributed across CPUs and people objects moved among locations, using various census, land-use and population-mobility data. It has been successfully applied to modelling of smallpox \cite{eubank2004modelling} and pandemic influenza plan \cite{halloran2008modeling}. The EpiSimdemics algorithm \cite{barrett2008episimdemics,bisset2012simulating} was based on contagion diffusion across a person-location graph, a bipartite graph where the nodes represented people and locations, while the edges indicated a person's presence at a location. A person-location graph of the entire population of the United States included 300 million nodes (people) and 1.5 billion edges, and ran on the 352,000 core NCSA BlueWaters system, so that 120 days of an epidemic were simulated in 12 seconds. While EpiSimdemics built the graph implicitly, the next extension, Epifast \cite{bisset2009epifast}, used an explicitly given network to simulate SEIR compartmental model in a distributed memory system with a master-slave computation, and run 10 times faster than EpiSimdemics for equivalent graphs.}

In order to inform policy at a national and state level, it is imperative for an epidemic model to have a high degree of spatial fidelity\new{, which is the primary motivation and focus of our work}. Due to the high concentration of the Australian population around urban areas, modelling Australian demographics provides a unique challenge. For instance, GLEAM~\cite{balcan2010modeling} is a global-scale simulation based on a gridded population of the world. In order to capture global disease dynamics, GLEAM discretises the world into a grid of 25$\times$25 km cells. While this method effectively captures global disease dynamics, it does not provide a resolution fine enough for modelling disease transmission at the individual level in Australia. A more recent simulator, FluTE~\cite{chao2010flute}, facilitates modelling epidemic scenarios in the United States but {is not immediately suitable to address} the Australian population, mobility, and school structure.

Agent-based simulators rely on detailed models of the transmission {and the natural history of the pathogen through a host individual~\cite{Longini2005}.}
Recent work by Cauchemez et. al.~\cite{cauchemez2011role} emphasizes the importance of high precision disease parameters when high fidelity census data is being used.
The authors recognised ``a lack of detailed data on transmission outside the household setting'' and quantified the parameters of disease spread within the school and home environment by observing an outbreak of H1N1(2009) in an elementary school in Pennsylvania.
Longini et al.~\cite{Longini2005} provided a detailed description of the disease natural history for influenza, specifying the progression of a disease from pathological onset to its eventual resolution within a host. There is significant research into estimating the proportion of individuals who become symptomatic (and hence more infectious)~\cite{carrat2008time,hayward2014comparative,leung2015fraction,feng2015technical}, informing our study.

\section{Simulation model} \label{sec:methodology}

\begin{figure}
  \centering
  \includegraphics[width=.8\textwidth]{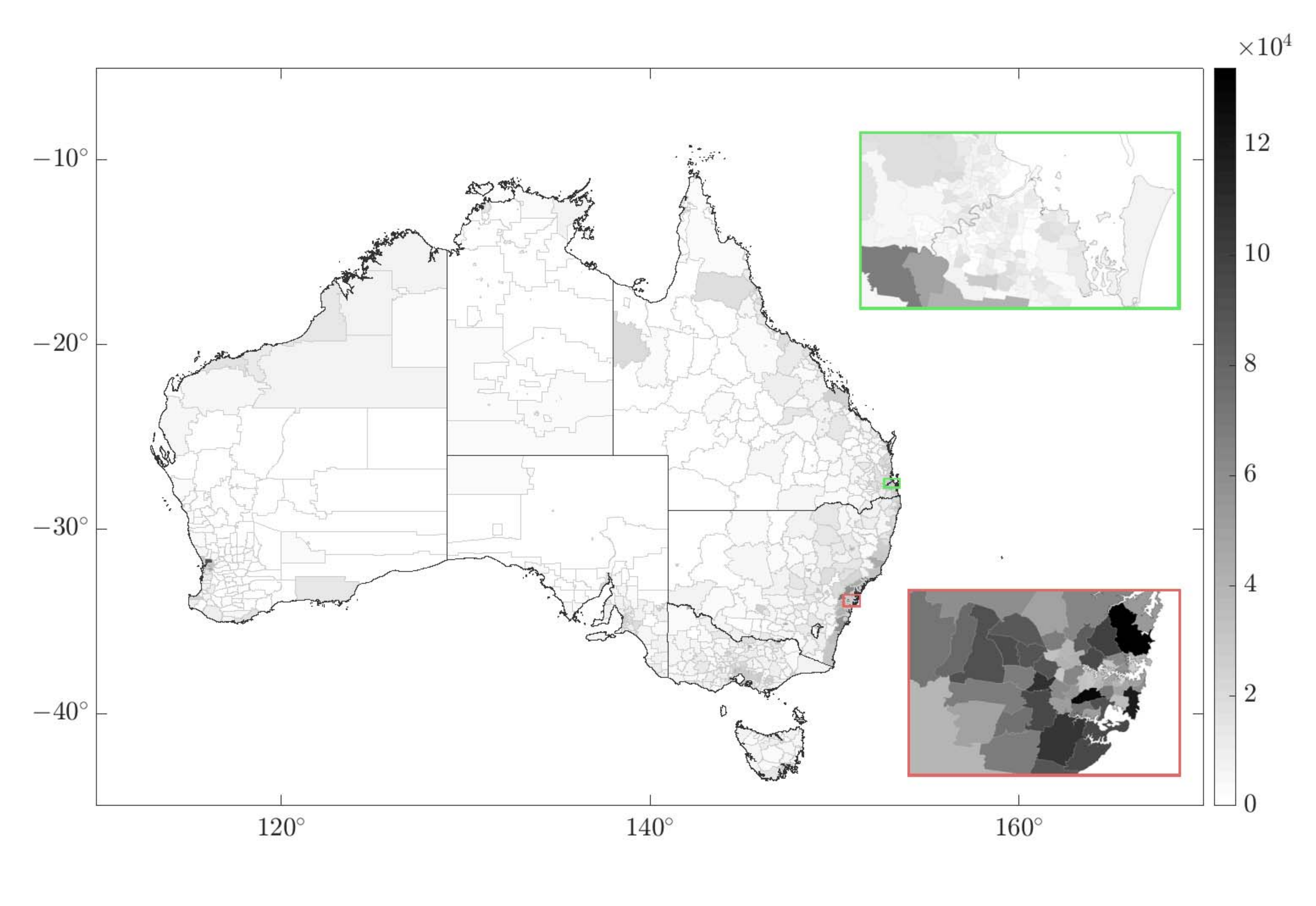}
  \caption{{Spatial distribution of the population of 19.8 million Australian residents according to 2006 Australian census data. The states and territories of Australia are (in order of population size): New South Wales (NSW), Victoria (VIC), Queensland (QLD), Western Australia (WA), South Australia (SA), Tasmania (TAS), Australian Capital Territory (ACT), Northern Territory (NT). Insets: distributions for \acp{SLA} based in and around Sydney (red) and Brisbane (green) on the same scale.}}
  \label{fig:demographics}
\end{figure}

Our simulation model comprises two basic elements: (i) a surrogate Australian population based on real-world data for demographics and worker flow; and (ii) a stochastic agent-based model for disease spread. The demographics induce numerous statistical areas (social \emph{mixing groups}) within which individuals associate and provide context for susceptible agents to contract an infectious disease. The model runs in cycles of two 12-hour periods (``day'' and ``night'') and an agent is associated with a set of daytime and nighttime groups depending on the cycle. Given the census data, we are able to generate mobility networks via the reported worker flow between these mutually exclusive sets. The framework of developing contact networks from census data and integrating a transmission model is extensively used in literature to stochastically model influenza pandemics~\cite{Longini2004,Longini2005,GermannKadauEtAl2006,chao2010flute,balcan2010modeling}.

\subsection{Demographics and mobility} \label{sec:demographics}


We utilise the 2006 Australian census data to distribute the model population of 19.8 million individuals into 1422 statistical local areas (\acp{SLA}). Figure~\ref{fig:demographics} shows the spatial distribution of the Australian population among the \acp{SLA}. These areas are further subdivided into \acp{CD} of approximately 225 dwellings each in metropolitan settings. The census provides population (e.g., age, sex, employment status) and housing (e.g., household size and composition) statistics at this geographic level. This data is then used to stochastically generate all households and individuals within each \ac{CD}.

Each individual in the population is a member of several mixing groups derived from census data and prior work~\cite{Longini2005,GermannKadauEtAl2006,chao2010flute}. The contact (and transmission) probability of any disease is, in general, decreasing with the group size, and numerous empirical studies report on these statistics for different contexts~\cite{Longini2005,GermannKadauEtAl2006,cauchemez2011role,chao2010flute}. During the daytime, the mixing groups comprise work groups for adults, and classrooms, grades, and schools for students. During the nighttime, individuals interact in households, household clusters, communities, and neighbourhoods.

We first stochastically generate the population based on housing statistics and then subsequently assign social mixing groups by geographic locality. Specifically, the census contains the frequency distribution of household sizes in each \ac{CD}. In addition, households are composed of either lone, group, or family types, and the distribution of these compositions are known for each \ac{CD}. {We further subdivide the family type} into the following categories: single (S), couples with children (CWC), couples without children (CWOC), {or single parent families (SPF)}. This yields a conditional distribution for family compositions, given household size, in each \ac{CD}. We then generate agents for the adults and children living in every residence by sampling a family type conditioned on the household size. Households have the highest pairwise contact probability and four consecutive houses are assigned to a larger social group denoted a cluster. The communities (\acp{CD}) and neighbourhoods (\acp{SLA}) within which these residents live then provide context for unspecified and occasional person-to-person contact (e.g., supermarkets, theatres, etc.). \new{In this study, the  population is stochastically generated once, and used in all simulations. It is possible to run simulations over multiple stochastically generated populations; however, this did not produce significantly different results for the key variables investigated in this work.} 

In the daytime, most working-age adults interact with other individuals in their immediate workplace, whereas students associate through local schools, grades, and classes. To establish working communities, the Australian census data subdivides \acp{SLA} into \acp{DZN}.
Although these geographic areas do not concord with \acp{CD}, the worker flow from \acp{CD} to \acp{DZN} can be used to accurately capture the short- to medium-distance population mobility important for disease spread~\cite{Longini2005}. Adults from each \ac{CD} are randomly assigned a \ac{DZN} based on this data, and the workers in each \ac{DZN} are split into working groups of around 20 adults to represent the colleagues with whom an individual is likely to work with on a daily basis. The commute distance distribution of workers given by this approach is shown {in the appendix in Fig.~\ref{fig:worker-commute}} and is distributed similarly to reports by the \ac{BITRE}~\cite{bitre15a}.

Finally, accurate contact networks in schools, grades, and classes are paramount to the demographics given the close and frequent contact of students with one another~\cite{cauchemez2011role}. Unfortunately, the census does not contain information about school attendance, nor spatial distribution of students. However, the \ac{ABS} provides state-wide frequency distribution tables specifying the number of schools with (non-uniform) intervals of the school attendance. Due to the concentration of the Australian population around urban areas, we employ a biased sampling technique to place these schools within \acp{DZN} that have more students nearby, i.e., \acp{SLA} that are more populated are biased to have more schools (see~\ref{app:schools}). Using this approach, the average commute distance of students is illustrated in Fig.~\ref{fig:student-commute} and is distributed according to reports by the \ac{BITRE}~\cite{bitre15a}.

%
%


\subsection{Transmission model} \label{sec:transmission-model}


At an individual level, we are concerned with the transition probability of a susceptible agent $i$ becoming infected during a given time period (step) $n$. Let the random variable $X_i(n)$ denote the state of the individual at time $n$. That is, we are interested in the infection transition probability:
\begin{equation}
	p_i(n) = \Pr\left \{ X_i( n ) = \textsc{latent} \mid X_i(n-1) = \textsc{susceptible} \right \},
\end{equation}
where \textsc{latent} is the state denoting exposure to the disease to that agent (see Sec.~\ref{sec:natural-history-model}). Section~\ref{sec:demographics} defined a number of mixing groups $\mathcal{G}_i(n)$ that each agent $i$ interacts within where each group $g\in \mathcal{G}_i(n)$ is associated with a static set of agents $\mathcal{A}_g$. The infection probability is thus computed as a product of all of the possible infectious contacts during that step:
\begin{equation} \label{eq:transition-prob}
	p_i(n) = 1 - \prod_{g \in \mathcal{G}_i(n)} \Bigg[ \prod_{j \in \mathcal{A}_g \setminus i} \left( 1 - p_{j \to i}^g(n) \right) \Bigg],
\end{equation}
where $p_{j \to i}^g(n)$ denotes the instantaneous probability of transmission from agent $j$ to agent $i$ in contact group $g$. Let $n_j$ denote the time agent $j$ becomes infected, then
\begin{equation} \label{eq:prob-transmission}
	p_{j \to i}^g(n ) = \kappa \ f( n - n_j \mid j, i ) \ q_{j \to i}^g.
\end{equation}
The scaling factor $\kappa$ is a free parameter that allows variation in the contagiousness of the simulated epidemic, while keeping other parameters constant.  Changing this parameter thus results in different values of the reproductive ratio $R_0$; we investigate the precise dependence between these two variables in Sec.~\ref{sec:r0-curve}. The function $f : \mathbb{N} \to [0,1]$ characterises infectivity of case $j$ over time; importantly, $f( n - n_j \mid j, i ) = 0$ when $n < n_j$. In this paper we assume infectivity linearly decreases over time. Finally, $q_{j \to i}^g$ is the probability of transmission from agent $j$ to $i$ at peak infectivity, which we discuss below. Each susceptible individual can become infected at the end of each step, which is determined by Bernoulli trial with transition probability $p_i(n)$. \new{Note that the probabilities of transmission $q_{j \to i}^g$ are very low in practice, and so the instantaneous probabilities of transmission $p_{j \to i}^g(n)$, scaled by $\kappa$, remain significantly below 1.}

\begin{figure}
    \subfigure[Natural history of the disease.]{ { \includegraphics[width=.46\textwidth]{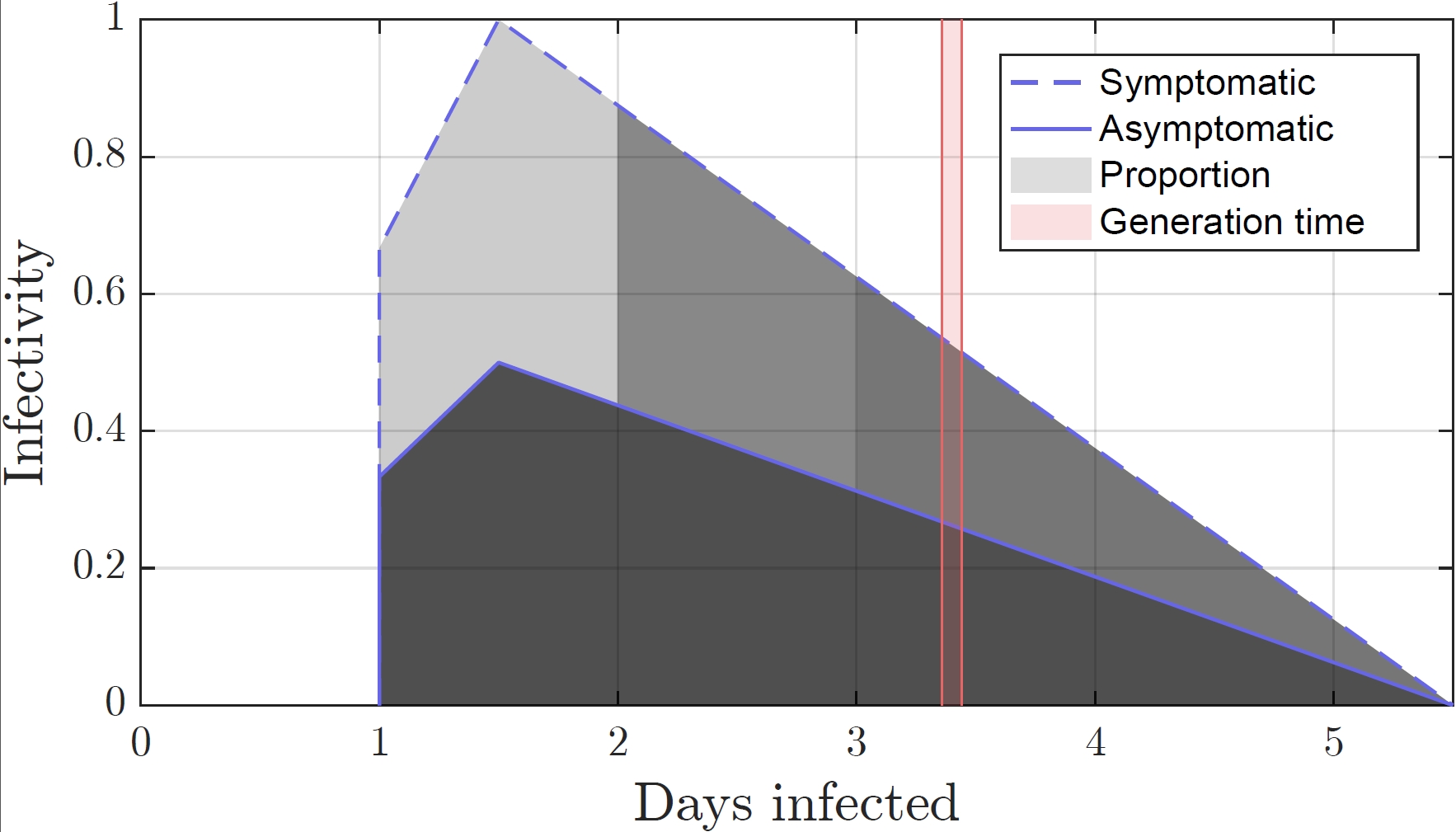} } \label{fig:infectivity} } \hfill
    \subfigure[{Simulated} generation time.]{ { \includegraphics[width=.47\textwidth]{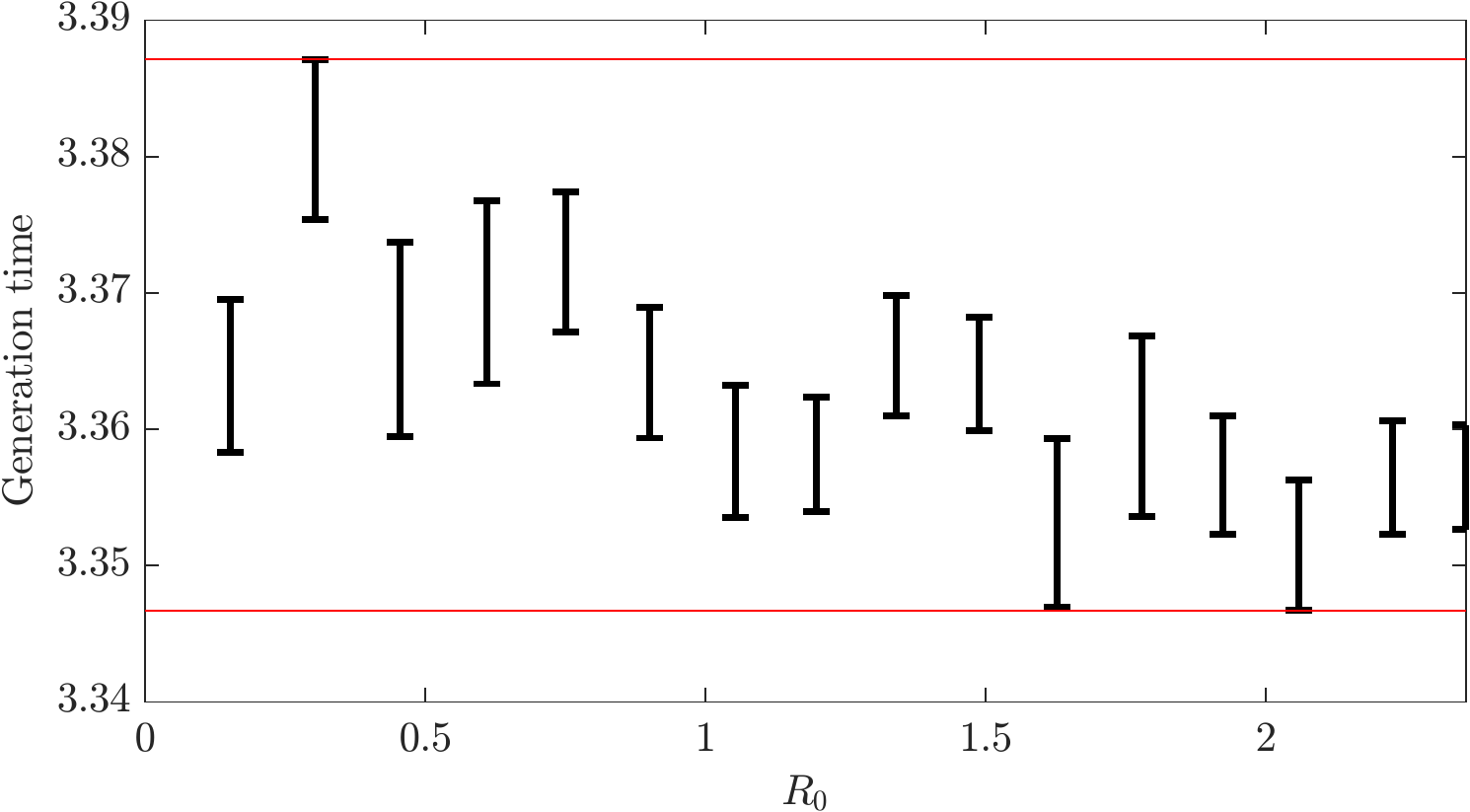} } \label{fig:generation-time} } \\
  \caption{Natural history of the disease and corresponding {simulated} generation time. The disease dynamics are modelled as having a linear increase followed by a linear decrease, as illustrated in~\ref{fig:infectivity}. In the figure, the area under the curve is shaded according to the proportion of people \emph{at least} that infectious after disease onset (darker representing a higher proportion). If an agent becomes symptomatic, their infectiousness doubles (dashed blue line) from that day onward. Moreover, 67\% of agents become symptomatic; of these agents, 30\% start showing symptoms on day 1, 50\% on day 2, and the remaining 20\% on day 3. {We obtain empirical generation times from simulations} resulting from this model, shown in~\ref{fig:generation-time} for a number of $R_0$ values. The confidence intervals range from 3.35 to 3.39 days (also shown on Fig.~\ref{fig:infectivity}), depending on $R_0$ and, in general, the generation time {has a slight downward trend as a function of disease severity.}} \label{fig:natural-history}
\end{figure}

In this work, we derive $q^g_{j\to i}$ from two sources: (i) the recent study of H1N1 (2009) by Cauchemez et al.~\cite{cauchemez2011role}, in setting transmission probabilities for a number of relevant mixing groups (households and schools), as well as (ii) a recent study using contact probabilities for mixing groups~\cite{chao2010flute}. The latter approach included mixing groups that were not considered by~\cite{cauchemez2011role} (household clusters, neighbourhoods and working groups). Furthermore, as pointed out by Chao et al.~\cite{chao2010flute}, these contact probabilities are in agreement with other agent-based models~\cite{halloran2008modeling,mossong2008social}. Thus, the transmission between two agents $q_{j \to i}^g$ within each contact group $g$ is described by either a transmission probability $\beta^g_{j \to i}$ or a contact probability $c_{j \to i}^g$.

The transmission probabilities contained in Table \ref{transmission_table} are obtained from a study of H1N1 (2009) by Cauchemez et al.~\cite{cauchemez2011role} which determined the transition rates of infection $\beta^g_{j \to i}$ between an individual $j$ and a susceptible individual $i$, using reverse jump Monte Carlo sampling methods, while fitting of the transition rates varied slightly depending on the mixing context $j$. We transform the rate of infection $\beta^g_{j\to i}$ into the daily probability of transmission from an infected individual $j$ to a susceptible individual $i$ as follows~\cite{cauchemez2011role} 
\begin{equation} \label{eq:cauch}
	q_{j \to i}^g = 1 - \exp \left( \beta^g_{ j \to i } \right).
\end{equation}

In addition, when the transition rates are not available in the study Cauchemez et al.~\cite{cauchemez2011role} we utilise the contact probabilities given in Table~\ref{tab:contact-table}, reported by~\cite{chao2010flute}. These probabilities represent the likelihood, within each daily period, of having a contact of sufficient duration and closeness for a transmission of the infection to be possible between the two individuals $i$ and $j$ in this social setting. Each contact probability $c^g_{j \to i}$ for a mixing group $g$ is multiplied by a factor $\rho$ scaling to the probability of transmission:
\begin{equation} \label{eq:ptrans}
	q_{j \to i}^g = \rho \ c^g_{j \to i}.
\end{equation}
The factor $\rho$ allows us to calibrate the transmission probabilities set by Eq.~\eqref{eq:ptrans} within mixing groups. This is achieved by matching the  attack rates \new{of the corresponding mixing groups} to the  values of \new{the context-dependent rates reported} in~\cite{cauchemez2011role}. This ensures consistency between Eq.~\eqref{eq:cauch} and Eq.~\eqref{eq:ptrans}.


\subsection{Natural history of disease} \label{sec:natural-history-model}

The natural history model describes the course of a disease from onset to resolution~\cite{Longini2005}. Each state has associated properties (e.g., infectious or non-infectious) depending on the disease being modelled. In this paper, we define a few states for influenza: \textsc{susceptible},  \textsc{latent}, infectious \textsc{symptomatic}, infectious \textsc{asymptomatic}, and \textsc{recovered}. Similar models have been established in earlier work based on observations of influenza dynamics~\cite{Longini2005}.

The model, illustrated in Fig.~\ref{fig:infectivity}, includes states that account for nuances in influenza that make the pathogen highly transmissible. The \textsc{latent} period captures the time from the exposure to the acquisition of infectiousness, contributing to the spread of the disease. From the \textsc{latent} state, 67\% of individuals develop symptoms (i.e., transition to	 \textsc{symptomatic})~\cite{carrat2008time}, leaving the remainder as \textsc{asymptomatic} agents that are half as infectious~\cite{Longini2005}. More recent studies of the proportion of symptomatic cases suggest a wider range of estimates -- within 25--67\%~\cite{hayward2014comparative,feng2015technical} -- but, in general, {this depends on} the study design and definitions of infection and symptomatic illness (with {estimates for asymptomatic cases ranging between} 0--100\%)~\cite{leung2015fraction}. Hence, we adopted the established estimates of~\cite{carrat2008time} that have been used in various simulations~\cite{GermannKadauEtAl2006,yang2009transmissibility,chao2010flute}.

An individual that will eventually show signs of illness will typically do so a number of days after contracting the disease. During the interim, these agents are considered as infectious as agents in the \textsc{asymptomatic} state. Specifically, out of the individuals that will become ill, 30\% will start showing symptoms {one} day after disease onset, 50\% after {two} days, and the remaining 20\% on day {three}. These dynamics result in the generation times (i.e., the times between the infection of a primary case and one of its secondary cases) between 3.35 and 3.39 days, depending on the $R_0$, as illustrated in Fig.~\ref{fig:generation-time}. These results are in line with the literature on generation times in a number of empirical studies~\cite{Ferguson} and established simulations~\cite{Longini2005,GermannKadauEtAl2006,chao2010flute}. In this framework, the prevalence at a given time is the number of \emph{ill} individuals in the \textsc{symptomatic} state. This is then used to compute the illness attack rate, $R_0$, and many other quantities of interest.

\subsection{Scenario description}

\begin{figure}
  \begin{minipage}{0.55\textwidth}
\bgroup
\setlength\arrayrulewidth{1pt}
\def\arraystretch{1.3}
\setlength\tabcolsep{4mm}
\rowcolors{1}{white}{gray!5!}
  \caption{Daily incoming passengers per Australian international airport obtained from \ac{BITRE}~\cite{BITREairport} along with a map detailing the airport locations.}
  \label{tab:airport}
        {\footnotesize
        \begin{tabular}{ l l l r }
          Airport code & State & City & Passengers \\
          \hline
          SYD & NSW & Sydney & 40884 \\
          MEL & VIC & Melbourne & 25859 \\ 
          BNE & QLD & Brisbane & 14250 \\
          PER & WA & Perth & 11449 \\
          OOL & QLD & Gold Coast & 3022 \\
          ADL & SA & Adelaide & 2214 \\
          CNS & QLD & Cairns & 1874 \\
          DRW & NT & Darwin & 597 \\
          TSV & QLD & Townsville & 105 \\ \hline
        \end{tabular}
    }
    \egroup
  \end{minipage}
  \begin{minipage}{0.32\textwidth}
    \includegraphics[width=\textwidth]{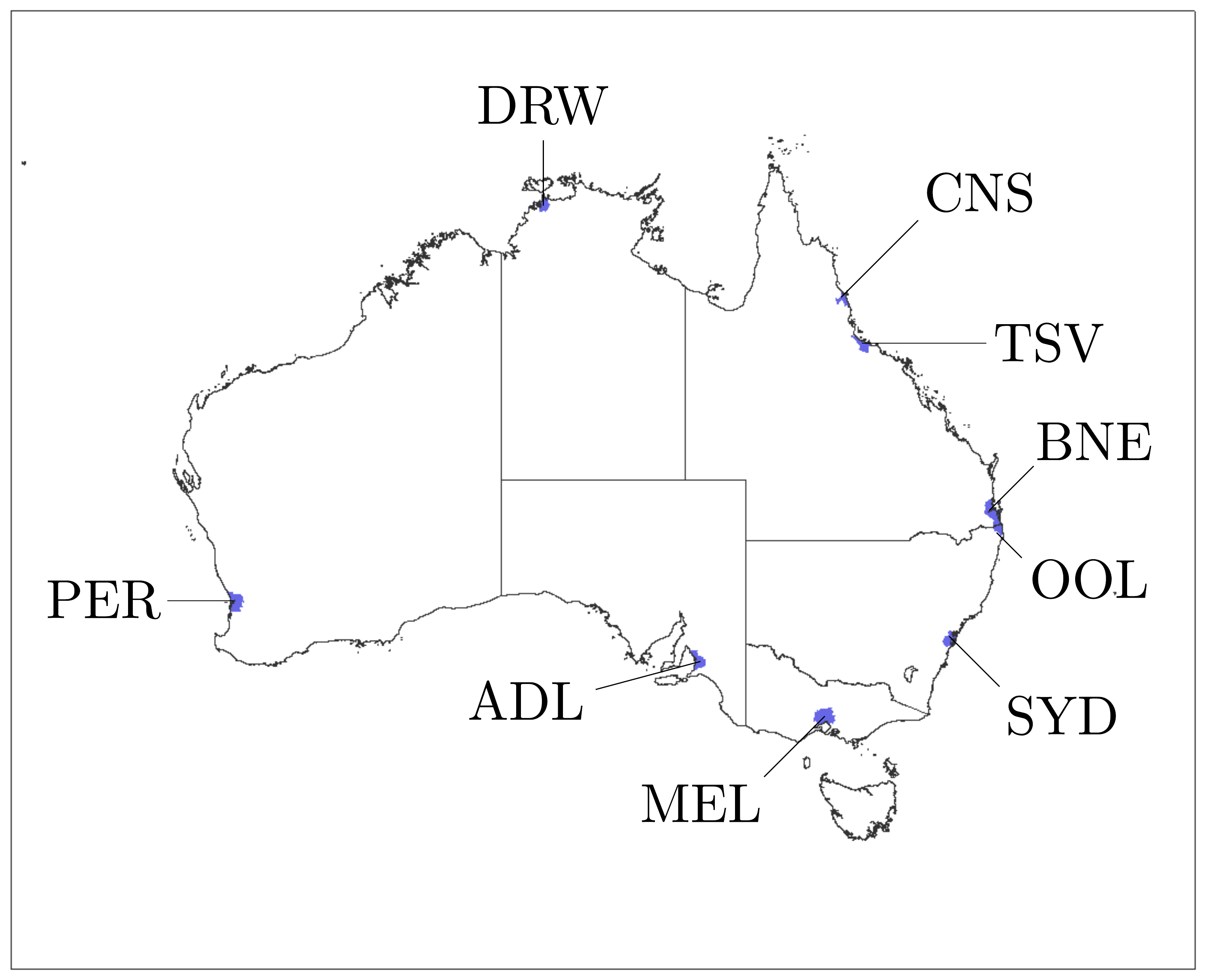}
  \end{minipage}
\end{figure}

In order to sufficiently model the domestic spread of the disease, we assume that the Australian population is exposed to the strain once it is a global pandemic. The geographic isolation of Australia suggests that people arriving from all other countries are equally likely to be already carrying the disease. Hence, we follow the approach of~\cite{GermannKadauEtAl2006} by introducing the disease at international airports, where the number infected is proportional to the air traffic inflow, obtained from \ac{BITRE}~\cite{BITREairport}.

The \acemod simulation itself is dynamically seeded such that new individuals are randomly infected at every time step $n$ instead of only at the beginning. There are 10 international airports in Australia and the number of ``seeds'' surrounding each airport is proportional to the average daily number of incoming passengers at that airport (see Tab.~\ref{tab:airport}). Note that the simulation is a closed system such that only existing and susceptible agents can be infected. Moreover, only individuals living in \acp{SLA} with geographic centroids inside a given (seeding) radius of each airport are considered for infection. The number of agents infected within this region are binomially distributed $B(P,N)$ with $P$ and $N$ chosen such that, on average, $0.04\%$ of incoming arrivals infect exactly one agent {(in agreement with the previous studies of~\cite{GermannKadauEtAl2006})}.

The epidemic characteristics are relatively insensitive to variations in the initial seeding, as has been shown before by~\cite{GermannKadauEtAl2006} and further confirmed by our study. However, the outbreak scenario does affect community synchrony, which we explore in Sec.~\ref{sec:synchrony-etc}.

\subsection{Calibrating the model}

Correct implementation of the transmission model requires calibrating a number of free parameters to known empirical results. That is, by rescaling the contact probabilities and transmission probabilities, we are able to match previously reported attack rates in schools, households, and other social mixing groups.

The first problem with calibrating the model is that contact groups all vary in size. The contact probabilities, reported in Tab.~\ref{tab:contact-table}, are determined for communities of fixed size following the study of~\cite{GermannKadauEtAl2006}. However, depending on the size of the community, these probabilities should be rescaled such that in larger communities people are less likely to be in contact with \new{a lower proportion of the population}. We do so by linearly rescaling these probabilities based on the number of agents in each stochastically generated group.

The second issue we address is calibrating the likelihood that a susceptible individual will contract a disease, given contact with another infectious individual, i.e., tuning $\rho$ {from Eq.~\eqref{eq:ptrans}}. This is done by matching the known proportion of transmissions within each different social group. That is, the fractions of influenza transmissions within households, school, and other settings (work, communities, etc.) have been estimated as 30--40\%, 20\%, and 40-50\%, respectively~\cite{yang2009transmissibility,cauchemez2011role}. Using these calibration procedures, the illness attack rates within different mixing groups in our simulation were matched to these studies, validating the simulations.

\section{Simulation results} \label{sec:results}

We illustrate the use of the model by studying influenza pandemics in Australia for a variety of simulated pathogens. We quantify the transmissibility of a strain by the \emph{reproductive ratio} $R_0$ and study the epidemic curves of the disease for a number of different $R_0$ values. In addition to analysing the (national) epidemic curve, we study the spatial distribution of the pandemic by prevalence proportion choropleths and disease synchrony.

\subsection{Reproductive ratio and the epidemic curve} \label{sec:r0-curve}

\begin{figure}
  \centering
  \includegraphics[width=.6\textwidth]{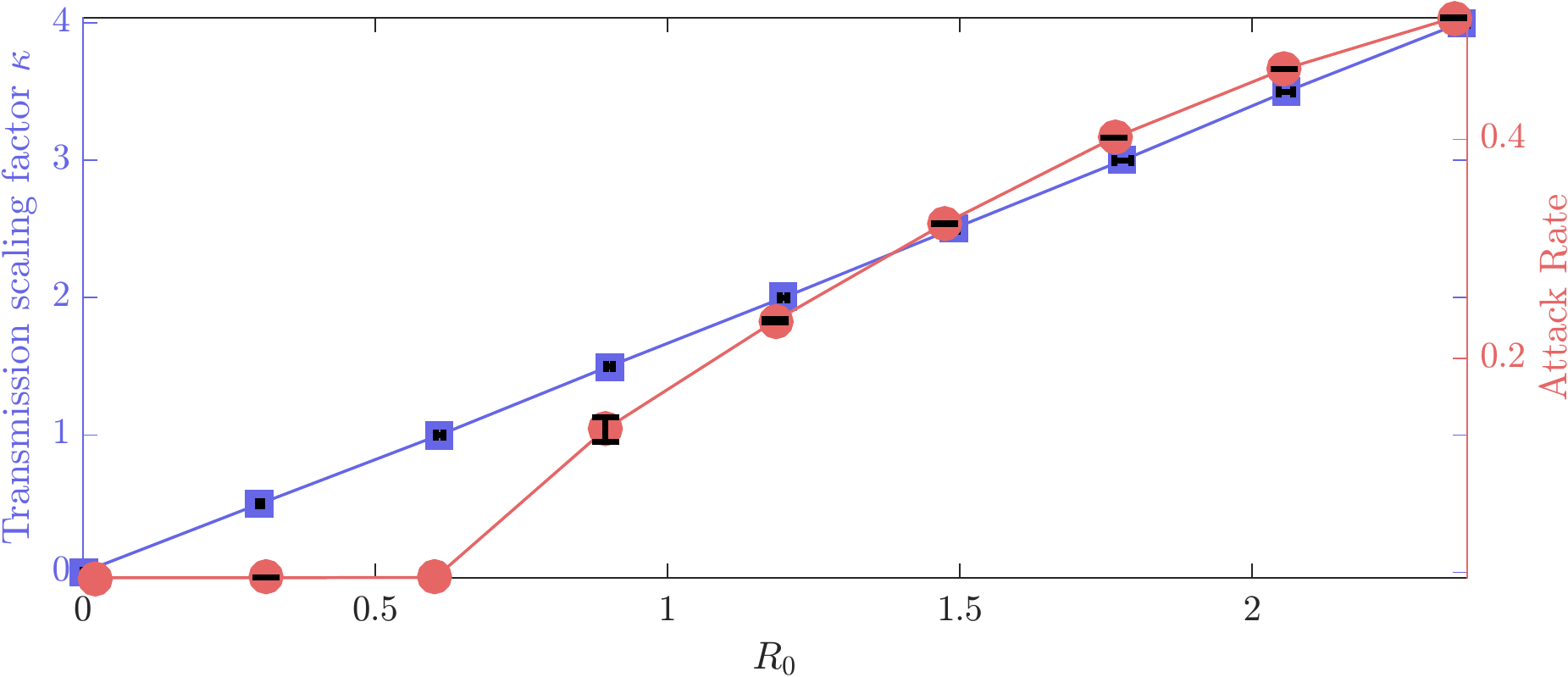}
  \caption{Reproductive ratio $R_0$, attack rate and transmission scaling factor $\kappa$ for simulated influenza epidemics in Australia.} \label{fig:r0-ar}
\end{figure}

{In the study of disease transmission, $R_0$ is defined by ``the expected number of secondary cases produced by a typical infected individual during its entire period of infectiousness in a completely susceptible population''~\cite{diekmann90a}. It is thus a key value that coarsely describes the rate of growth of an infection and whether it will spread in the population.}

We illustrate both analytically and empirically that we can approximate $R_0$ by a linear function of $\kappa$ for a biologically plausible range of values. A similar result was empirically shown by Chao et al.~\cite{chao2010flute}, who noted that $R_0$ is linear in $\kappa$ (from Eq.~\eqref{eq:prob-transmission}). Using Eq.~\eqref{eq:prob-transmission} and taking ``typical'' to mean ``in expectation'', we obtain the $R_0$ from Eq.~\eqref{eq:transition-prob}:
\begin{align} \label{eq:r0-solved}
	R_0 &= \mathbf{E} \Big( \sum_i X_i(N) = \textsc{symptomatic} \mid X_J(0) = \textsc{symptomatic} \Big) \nonumber \\
	    &= \mathbf{E} \Big( \mathbf{E} \Big( \prod_n p_I(n ) \mid X_J(0) = \textsc{symptomatic} \Big) \Big) \nonumber \\
		&= \sum_j \sum_i \prod_n \Big( 1 - \prod_{g \in \mathcal{G}_j(n)} \left( 1 - p_{j \to i}^g(n) \right) \Big),
\end{align}
where $\mathcal{G}_i(n) = \mathcal{G}_j(n)$ by the commutative property of the mixing groups. By expanding the inner and outer products in Eq.~\eqref{eq:r0-solved}, we note that the dominant terms are the first-order components $\sum_g p_{j \to i}^g(n) = \kappa \ \sum_g f(n - n_j \mid i,j) q_{j\to i}^g$. Thus, we expect an approximately linear relationship of $R_0$ for reasonable values of $\kappa$. In order to validate our conjecture, we compute the sample population $R_0$ by performing local simulations for a given set of parameters. Within each simulation, one agent is seeded at random, the number of cases directly caused by this agent is counted, then $R_0$ is computed as the average of this number. Using this approach, we evaluate $R_0$ for a number of $\kappa$ values, as shown in Fig.~\ref{fig:r0-ar}, where the result is a linear and homoscedastic function; the {simulation} results thus support our analytical reasoning. Figure~\ref{fig:r0-hists} in the appendix illustrates the frequency distribution of secondary cases for $\kappa = \{1.0, 2.0, 3.0, 4.0\}$.

For a given $R_0$, we measure the severity of the epidemic via the \emph{prevalence}, \emph{incidence}, and \emph{infection attack rate} of a disease. The prevalence and incidence are defined by the number of current and newly ill agents at each discrete time step. Both quantities are commonly used to characterise the time dependent global state of the population with respect to the epidemic~\cite{GermannKadauEtAl2006,chao2010flute,balcan2010modeling}. Tracing the incidence proportion {(normalised by population size)} over time produces an \emph{epidemic curve}. In addition, the infection attack rate captures the proportion of the population that became ill over the duration of the epidemic~\cite{porta2014dictionary} and is quantified by cumulating the incidence proportion. The attack rate is thus an effective summary measure for the time independent disease spread.

\begin{figure}[t]
  \includegraphics[width=.8\textwidth]{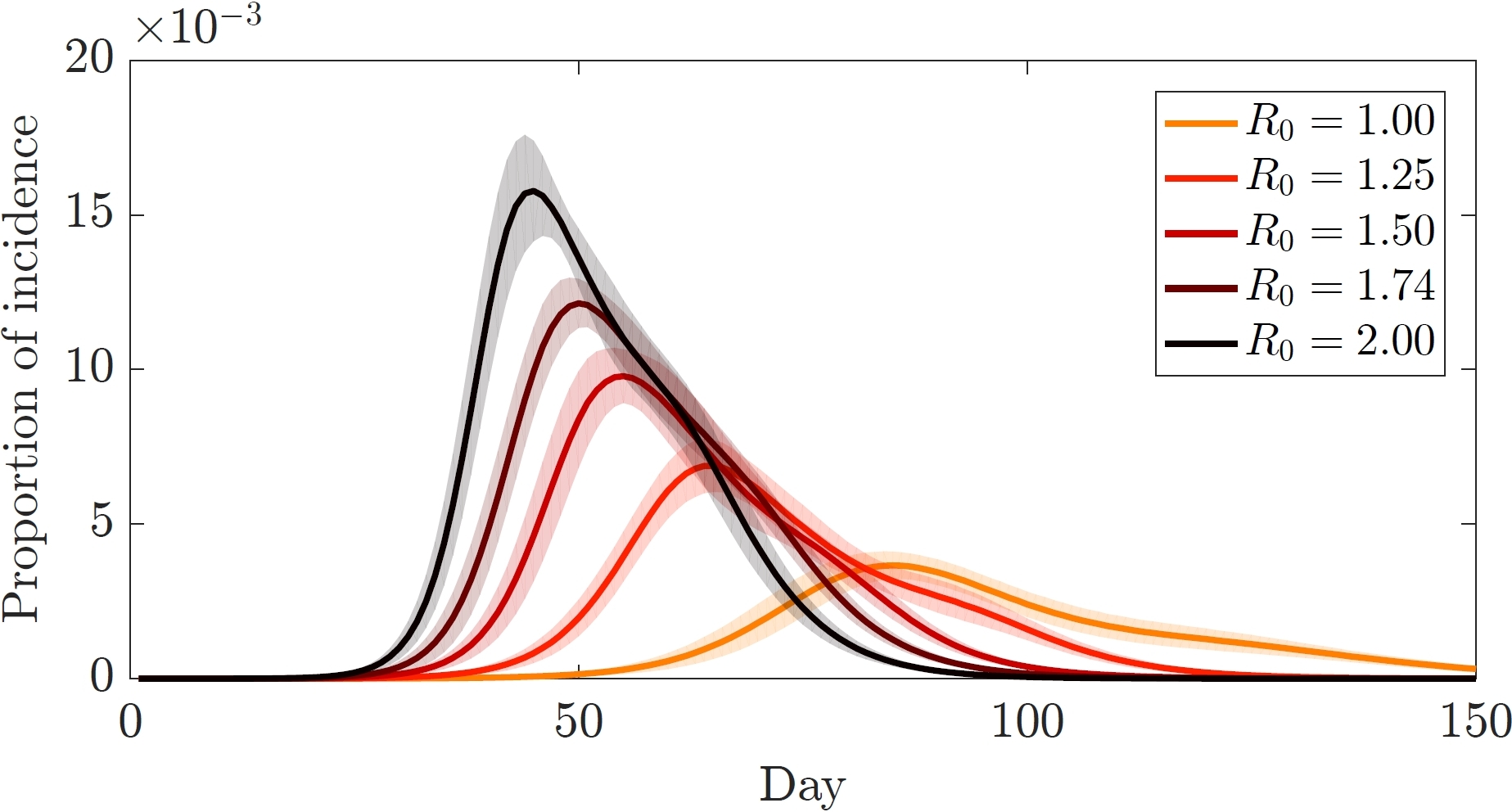} \label{fig:inc_curve}
  \caption{The epidemic curve illustrates the trajectory of the epidemic by tracing the incidence over time for each $R_0$ value. This illustrates the behaviour of the simulated influenza epidemics in Australia with $R_0 = \{1.0, 1.25, 1.5, 1.75, 2.0\}$. {We performed $10$ trial runs for each $R_0$ value, then plot the mean (solid lines) and standard deviation (shaded area).}} \label{fig:curve-table}
\end{figure}

Figure~\ref{fig:curve-table} illustrates the global epidemic state for simulated pathogens. As expected, a simulated strain with a higher $R_0$ is associated with higher attack rate and an earlier incidence peak. Table~\ref{tab:aggregated} gives corresponding key features of the incidence curve for the range of $R_0$ values used. {The reported illness attack rates match expected empirical and simulated results~\cite{GermannKadauEtAl2006,yang2009transmissibility,chao2010flute}, validating our model.}

Moreover, Tab.~\ref{tab:local} collates the national and the community average attack rates for the different age groups. {We compute the national rate} as the cumulative number of ill individuals in the country, giving a weighted average of the attack rate depending on the \ac{SLA} size.
{The community based measure is an unweighted average of each \ac{SLA} attack rate, obtained by taking the mean of the attack rate in that community.}
The table illustrates the significant impact on student-aged individuals, having approximately four times higher incidence than the other groups. Further, the community and national attack rates converge for all age groups for higher $R_0$, {indicating the attack rate saturates uniformly for all \acp{SLA}.} This is further evidenced by the synchrony increasing with higher $R_0$ in Tab.~\ref{tab:aggregated}.

\bgroup
\def\arraystretch{1.3}
\setlength\tabcolsep{4mm}
\rowcolors{1}{white}{gray!5!}
\begin{table}[t]
  \caption{Characteristics of the incidence curves{, averaged over 10 trials.}}
  \label{tab:aggregated}
  \resizebox{.8\textwidth}{!}{%
    {\raggedright
      \noindent
      \begin{tabular}{L{.45\textwidth} c c c c c }
        \hangpara{3ex}{1} Basic reproductive number $R_0$ & $1.0$ & $1.25$ & $1.5$ & $1.75$ & $2.0$ \\ 
        \hline
        \hangpara{3ex}{1} Rate of spread: 1K ill$^*$  & $29$ & $22$ & $21$ & $18$ & $17$ \\
        \hspace{3ex}10K ill$^*$ & $44$ & $33$ & $30$ & $25$ & $24$ \\
        \hspace{3ex}100K ill$^*$ & $60$ & $44$ & $39$ & $33$ & $31$ \\
        \hspace{3ex}1M ill$^*$ & $80$ & $58$ & $50$ & $42$ & $39$ \\
        \hangpara{3ex}{1} Peak of epidemic$^*$ & $84$ & $63$ & $59$ & $47$ & $47$ \\
        \hangpara{3ex}{1} Daily number of new cases at peak activity & $85.3$ K & $140$ K & $189$ K & $257$ K & $328$ K \\
        \hangpara{3ex}{1} Number of days with $>$ 100K ill & $44$ & $55$ & $52$ & $51$ & $48$ \\
        \hangpara{3ex}{1} Cumulative number of ill individuals & $3.4$ M & $5.0$ M & $6.4$ M & $7.7$ M & $8.8$ M \\
        \hangpara{3ex}{1} Synchrony of community epidemic peaks ($10^{-3}$) & $1.38$ & $2.67$ & $4.33$ & $6.39$ & $7.70$ \\
        \hline
        \hangpara{3ex}{1} M -- Million; K -- Thousand. \\
        \cellcolor{white} \hangpara{3ex}{1} $^*$ Days after initial introduction.
      \end{tabular}
    }}
  \end{table}
  \egroup
  
  \bgroup
  \def\arraystretch{1.3}
  \setlength\tabcolsep{4mm}
  \rowcolors{1}{white}{gray!5!}
  \begin{table}
    \caption{Attack rates, per 10K individuals, at the community and national level for different age demographics{, averaged over 10 trials}. The attack rate is the proportion of the population which became ill. The community attack rate is calculated as the average of the attack rates of each SLA.}
    \label{tab:local}
    \resizebox{\textwidth}{!}{%
      {\raggedright
        \noindent
        \begin{tabular}{p{.35\textwidth}  l l  c c c c }
          & Age Group & $R_0 = 1.0$ & $R_0 = 1.25$ & $R_0 = 1.5$ & $R_0 = 1.75$ & $R_0 = 2.0$ \\
          \hline
          \cellcolor{gray!5!} & 0-4   & $1164$  & $1898$  & $2600$  & $3270$  & $3870$ \\
          \cellcolor{gray!5!} & 5-18   & $4140$  & $5080$  & $5560$  & $5860$  & $6070$ \\ 
          \cellcolor{gray!5!} Cumulative number of community & 19-34 & $967$  & $1698$  & $2500$  & $3250$  & $3970$ \\ 
          \cellcolor{gray!5!} \hspace{2ex} illnesses per 10K$^*$ & 35-64 & $975$  & $1710$  & $2500$  & $3250$  & $3950$ \\
          \cellcolor{gray!5!} & 65+   & $939$  & $1665$  & $2450$  & $3210$  & $3920$ \\ 
          \cellcolor{gray!5!} & Overall & $1552$  & $2320$  & $3050$  & $3710$  & $4320$ \\ 
          \hline
          \cellcolor{gray!5!} & 0-4 & $1280$  & $2020$  & $2730$  & $3370$  & $3960$ \\ 
          \cellcolor{gray!5!} & 5-18 & $4520$  & $5400$  & $5880$  & $6140$  & $6310$ \\
          \cellcolor{gray!5!} Cumulative number of national & 19-34 & $1042$  & $1774$  & $2590$  & $3320$  & $4010$ \\
          \cellcolor{gray!5!} \hspace{2ex} illnesses per 10K$^*$ & 35-64 & $1082$  & $1824$  & $2630$  & $3360$  & $4060$ \\ 
          \cellcolor{gray!5!} & 65+    & $1003$  & $1724$  & $2520$  & $3260$  & $3940$ \\
          \cellcolor{gray!5!} & Overall & $1734$  & $2500$  & $3230$  & $3870$  & $4450$ \\ 
          \hline				
          \cellcolor{white} \hangpara{4ex}{1} $^*$Compared to the number of agents in that age group (e.g., per 10K 19-34 year olds).
        \end{tabular}
      }
    }
  \end{table}
  \egroup

\subsection{Spatial spread of the disease} \label{sec:synchrony-etc}

\begin{figure}
  \subfigure[$R_0 = 1.5$]{{\includegraphics[height=190mm]{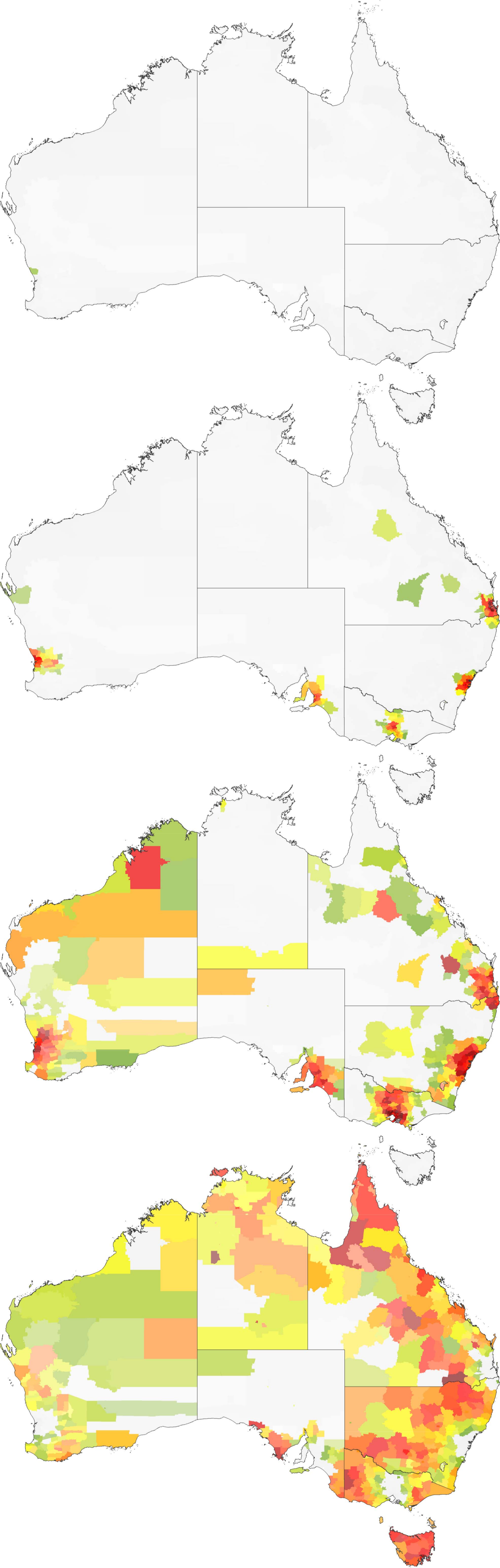}}\label{fig:p_sla_r1-75}} \hspace{12.5mm}
  \subfigure[$R_0 = 2.0$]{{\includegraphics[height=190mm]{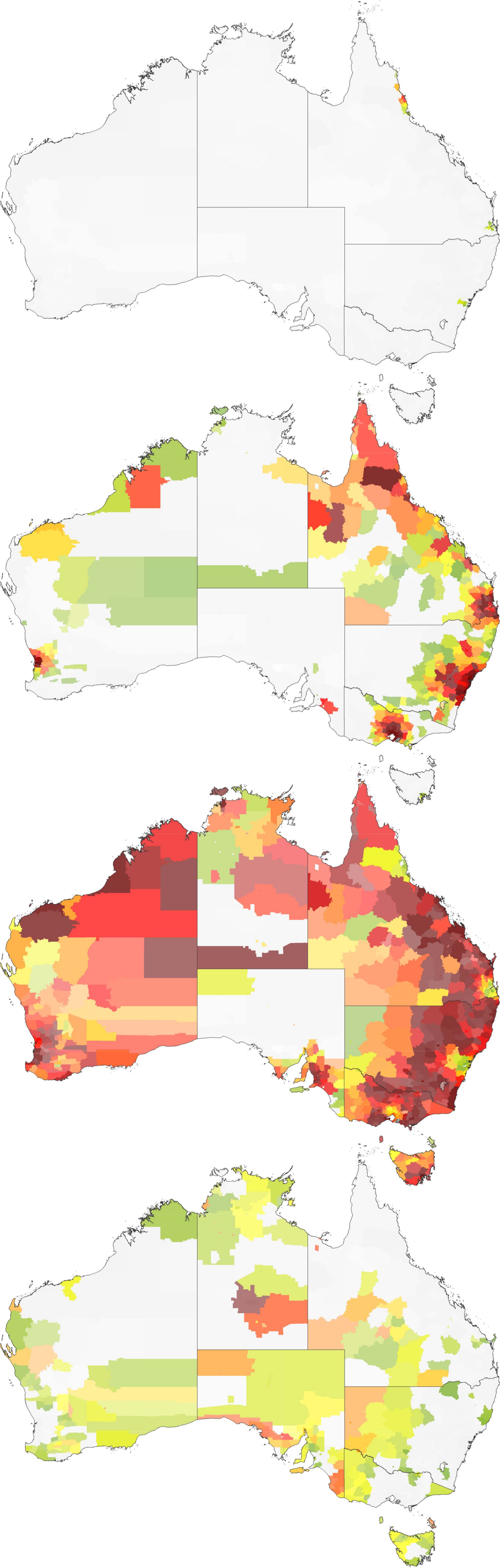}}\label{fig:p_sla_r2}} \hspace{5mm}
  \subfigure{{\includegraphics[height=50mm]{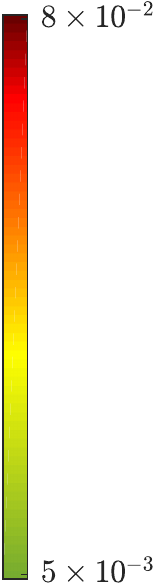}}\label{fig:cbar}} \hfill\null
  \caption{Prevalence proportion choropleths illustrating the spatial distribution of simulated influenza pandemics in Australia for $R_0 = 1.5$ and $R_0 = 2$. For the colour scale, the minimum prevalence (green) is $5 \times 10^{-3}$ and maximum prevalence (red) is $8 \times 10^{-2}$. We plot the distribution for days 30, 50, 62, and 88, with time going down the page. The pandemic spreads from these urban hubs inwards toward central Australia. Both simulations are sample realisations comprising the same demographics (contact) and mobility networks, as well as identical seeding	at the same rate at major international airports around Australia (see Tab.~\ref{tab:airport}). The epidemic peaks at larger cities at similar times, whereas less populous areas are less likely to synchronise.}\label{fig:spatial-aus}
\end{figure}

\begin{figure}
  \subfigure[]{\includegraphics[height=55mm]{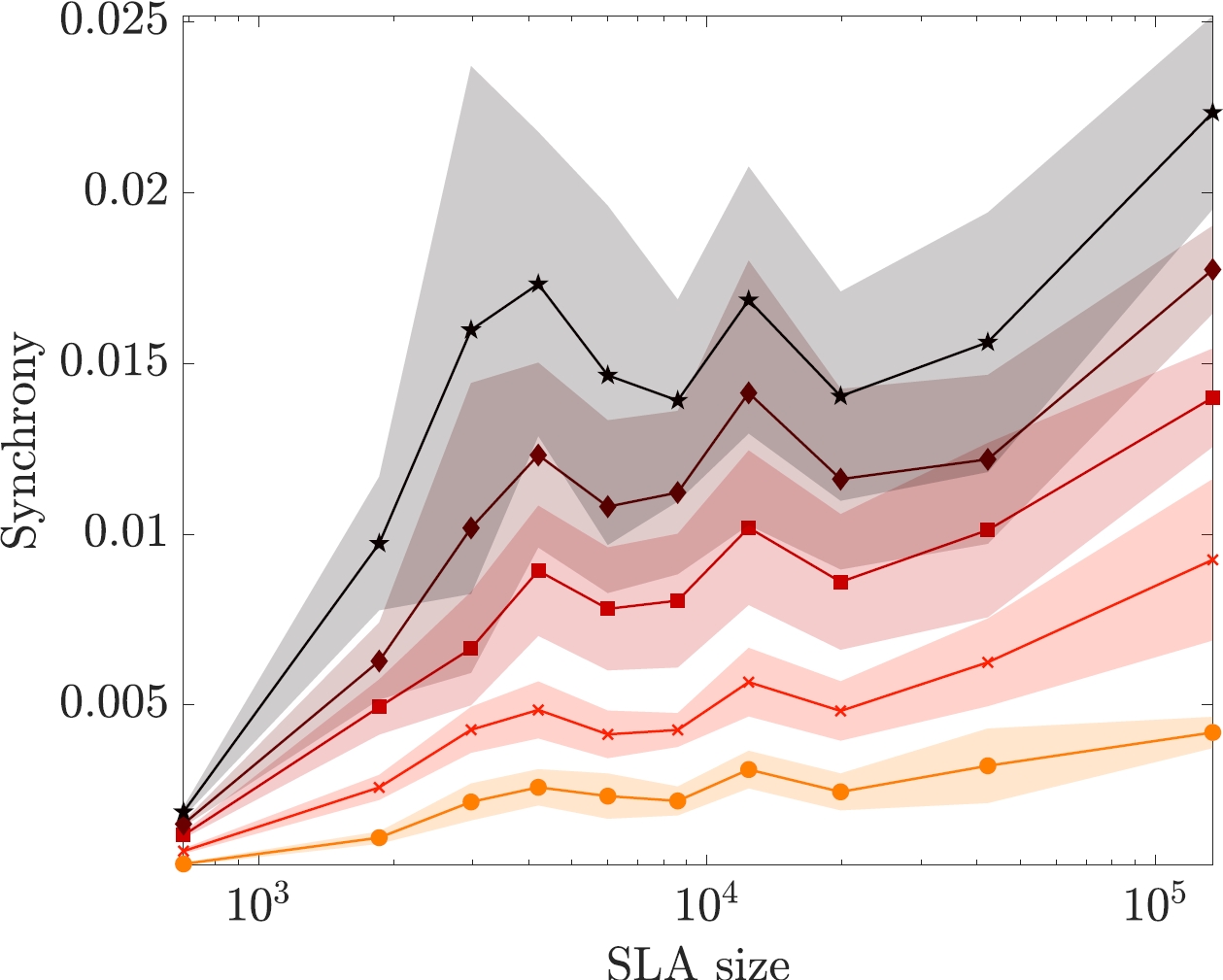} \label{fig:synchrony-a} } 
  \subfigure[]{\includegraphics[height=55mm]{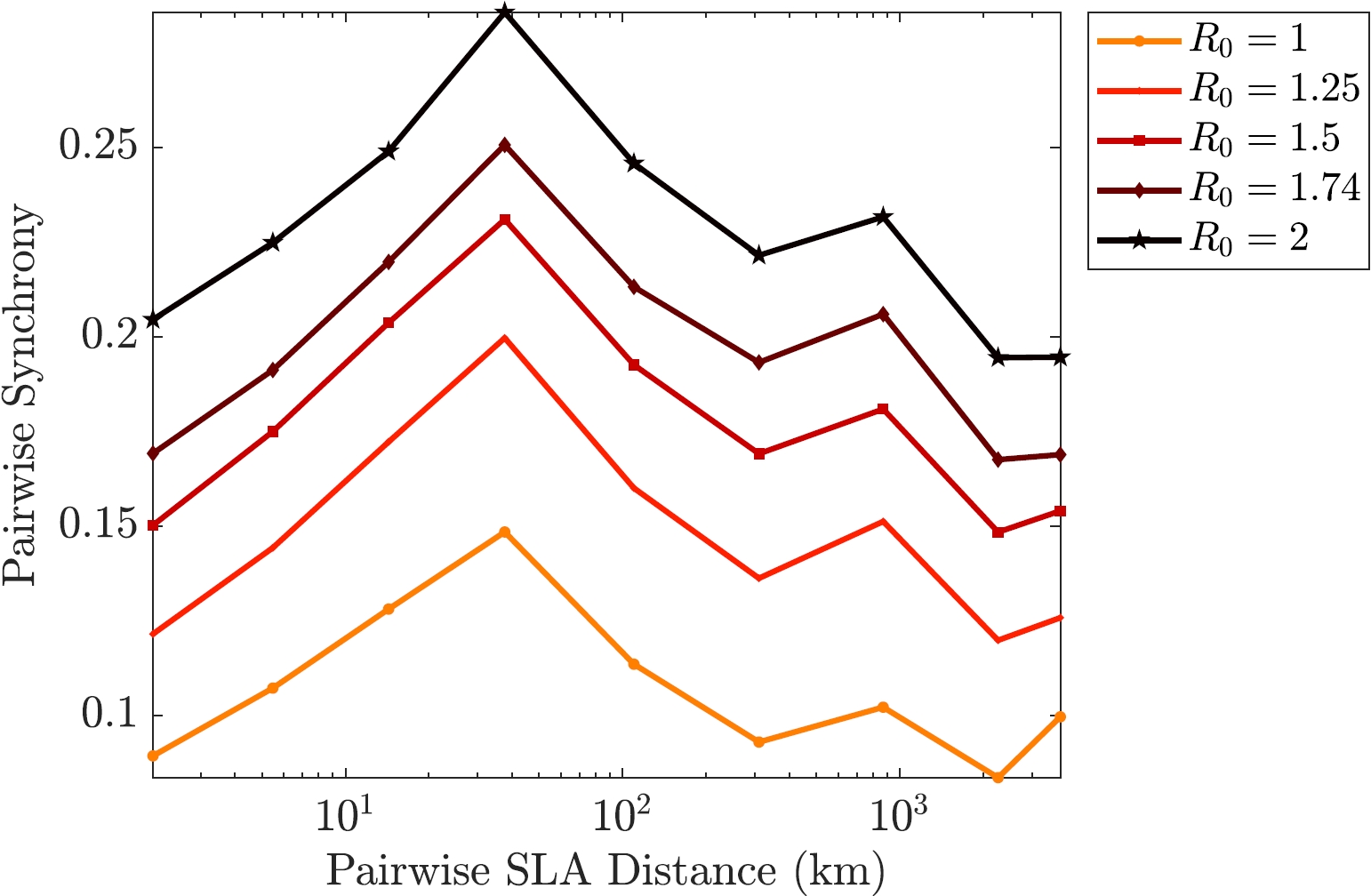} \label{fig:synchrony-b} } \\
  \subfigure[]{\includegraphics[width=\textwidth]{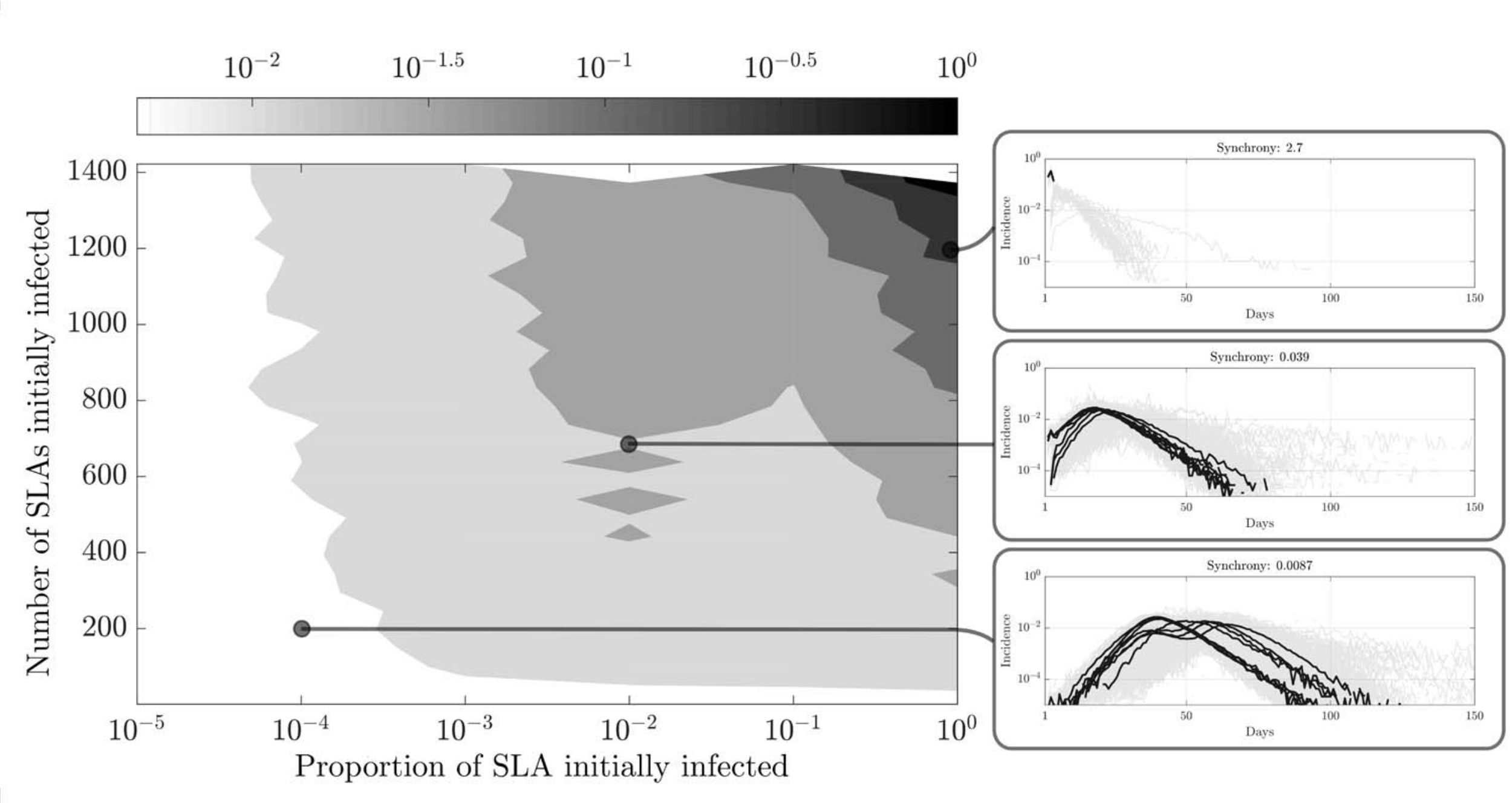} \label{fig:synchrony-examples} }
  \caption{Synchronisation of epidemic curves between the different communities (\acp{SLA}) in Australia illustrating spatial hierarchies. Figure~\ref{fig:synchrony-a} investigates how synchronisation is related to community (\ac{SLA}) size, i.e., communities of similar size are grouped together and their peak incidence times are compared. \new{Synchrony is computed for 10 partitions (bins), with around 142 SLAs per interval. 
		 Figure~\ref{fig:synchrony-b} traces the pairwise synchrony between all pairs of SLAs against the pairwise distance, averaged over 10 partitions (bins).} 
		Figure~\ref{fig:synchrony-examples} illustrates the effect of different {outbreak} scenarios on the \new{synchrony} of each community. This is achieved by seeding the pathogen at a varying number of \acp{SLA}, whilst varying the proportion of those \acp{SLA} \new{initially} infected. To the right of the figure we illustrate example attack rates (cumulative number of ill individuals) for given datapoints in the contour plot. The bold black lines illustrate the attack rate for the most populous \acp{SLA} in each state, whereas grey lines are all other \acp{SLA}. We observe an increased \new{synchrony} if the infection starts in a larger number of \acp{SLA} and with a higher proportion of infected individuals within these \acp{SLA}.} 
  \label{fig:synchrony-and-examples}
\end{figure}

We demonstrate the spatiotemporal spread of the simulated pandemics qualitatively and quantitatively through prevalence proportion choropleths and epidemic synchrony. Both tools are employed at the community (\ac{SLA}) level in order to identify important hubs and turning points of the developing epidemic.

In Fig.~\ref{fig:spatial-aus} we present the spatial distribution for a realisation of the simulated influenza pandemics for $R_0 = 1.5$ and $R_0 = 2$. The snapshots are taken from key days that were identified from the epidemic curve in Fig.~\ref{fig:curve-table}: onset inflexion point\footnote{The inflexion point is approximated to be between the two days.} (day 30); peak of $R_0 = 2.0$ (day 50); peak of $R_0 = 1.5$ (day 62); and intersection point (day 88). Due to the seeding of the pathogen at major airports and the contact network, the epidemic tends to spread primarily around the coast before it can be traced inland, where often entire \acp{SLA} are left unaffected in central Australia. Moreover, the incidence typically peaks first in dense urban communities, followed by sporadic outbreaks in the less populous areas. This further motivates quantifying this behaviour by analysing the synchrony of community-based epidemic curves. 

Epidemic synchrony is characterised by the phase coherence of incidence curves in specified geographic locations~\cite{viboud2006synchrony}. In this simulation, however, each \ac{SLA} follows a relatively smooth transition, allowing us to simplify the definition. We denote asynchrony of a set of communities by the variance of their incidence peaks (in days) and, consequently, synchrony is the reciprocal of this quantity. Figure~\ref{fig:synchrony-and-examples} gives results on the \new{synchrony} of incidence peaks in \acemod simulations. {It has been established that more populous regions of the United States exhibit more synchronised epidemics, i.e., that there exists a ``hierarchical spatial spread'' of influenza~\cite{viboud2006synchrony}. We investigate this hypothesis for Australian demographics (that are most prominently characterised by concentration of the population around urban and coastal regions). In Fig.~\ref{fig:synchrony-a} we group communities of similar size together and their synchrony is computed. 

\new{Using the synchrony measure, we study the phenomena of both hierarchical and wave-like spread. While the hierarchical spread is captured by the synchrony across SLAs of similar sizes, the wave-like spread is associated with pairwise geographical distance. That is, to investigate the hierarchical hypothesis, we partition all SLAs into groups according to their population size and calculate the synchrony across all peak times within each group. This is shown in Fig.~\ref{fig:synchrony-a}, where we see an increase in synchrony as the SLA size grows. In contrast, in order to verify the wave-like hypothesis, we compute pairwise synchrony between all pairs of SLAs, and trace it against the pairwise distance. Fig.~\ref{fig:synchrony-b} shows the average pairwise synchrony for ten partitions (bins) of  the pairwise distance. Crucially, the average pairwise synchrony does not grow with the increases in the pairwise distance. These results support the hypothesis of a hierarchical spatial spread, suggesting the regional movement of disease depends on the worker/student flow moreso than the geographic distance.

We also note that the synchrony with respect to SLA size, shown in Fig.~\ref{fig:synchrony-a}, is particularly pronounced for two specific SLA sizes. 
This two-phase characteristic is also evident in incidence plots (cf. Fig.~\ref{fig:inc_curve}), suggesting a secondary phase of an epidemic. The precise dependency between such phases and a multi-modal character of the synchrony profiles is a subject of future research.}

{Finally, we study the effect of outbreak scenarios on epidemic synchrony by varying the number of \acp{SLA} infected at the beginning of the simulation, as well as the proportion of individuals seeded within these \acp{SLA}.}
The number of \acp{SLA} initially infected ranges from 1 to 1422, and the proportion of each \acp{SLA} infected is logarithmically spaced between $10^{-5}$ and $1$.
{We then compare the synchronisation between community incidence peaks when we vary these two parameters, as represented with the contour plot shown in Fig.~\ref{fig:synchrony-examples}}.

\section{Discussion and future work} \label{sec:discussion-and-future-work}

In this paper we present the \acemod simulator as an approach for modelling epidemics in Australia using census data. Specifically, we used the 2006 Australian Census data to construct a contact network and develop a stochastic agent-based model for disease spread. The detailed data on population demographics and mobility allowed us to generate specific dynamics of influenza viral infection and transmission. {Moreover, the transmission and disease natural history models were calibrated to empirical data on influenza transmission, allowing us to validate the model. The agent-based model facilitated fine-grained analysis of} the spatiotemporal dynamics of influenza spread both qualitatively and quantitatively through incidence curves, prevalence choropleths, and epidemic synchrony.

\new{This study has highlighted several interesting dependencies in large-scale epidemics, calibrated for Australian demographics and connectivity. Firstly, we showed that the reproductive ratio $R_0$ can be approximated by a linear function of a simple factor that scales the probability of disease transmission. Secondly, the epidemic curves,  produced in our simulations, over a typical range of reproductive ratios, concur with the general ranges of the community and national attack rates, and show, in addition, the presence of secondary phases following the primary peaks. 
The spatiotemporal spread of the simulated pandemics was also illustrated using prevalence proportion choropleths, identifying important hubs and turning points of an epidemic. 
Most importantly, we verified the hypothesis that the modelled disease spreads in accordance with spatial hierarchies as opposed to a wave-like fashion. This has been achieved by measuring synchrony as a function of both population size and pairwise distance between SLAs.  As the number of  initially infected \acp{SLA} and the proportion of each infected \acp{SLA}  are increased, the synchrony between communities increases, suggesting that spatial hierarchies disappear depending on how widespread the pandemic already is when it reaches Australia. 

There are several important assumptions behind the presented results: fidelity of the Australian census datasets, including mobility patterns; knowledge of the transmission rates in specific mixing groups; and understanding of the natural history of the disease. With census datasets improving in their overall quality and coverage, there is an increasing confidence in data-driven simulation studies. In parallel, empirical epidemiological studies continue to narrow down the uncertainty in key modelling parameters. One of the emerging trends is thus an integrated real-time tracking of epidemic spreads with rapid simulation of ``what-if'' scenarios.}

It is known that infectious disease dynamics are heavily influenced by social network characteristics. It is then crucial to better represent and characterise the various Australia social networks in order to both ameliorate our predictions and suggest appropriate mitigation strategies. For this, we first aim to leverage the 2011 and 2016 census data to generate surrogate populations. With this, we could also study how the changes in the social network over time affect our predictions under the same disease dynamics. This could in turn help us in tuning our model to account for the predicted future changes in the social network structure.

Subsequently, our goal is to study the network characteristics to better understand the relationship between epidemic dynamics and different network properties. For instance, the critical value of the reproductive ratio $R_0$ (and its relationship to attack rate) is influenced by the network topology, and various graph-theoretic techniques can be employed to reveal these dependencies \cite{ PPZ2009,kitsak,PPZ2012,daraganova2012networks,PPH13}.

Directed information-theoretic measures, such as transfer entropy~\cite{schreiber-2000-85}, are particularly suited to this task in real systems in either the time-averaged or local forms~\cite{liz01}. These quantities have been used previously to infer interaction networks in multi-agent systems~\cite{cliff2017quantifying} and are particularly suited to processes with attracting states~\cite{cliff2016information,cliff2018minimising}. This would allow us to study contact networks in terms of both structural and functional connectivity \cite{ fris94,hon07,bett07,pradhana} and their dominant motifs \cite{Milo,spo04,PPZ2008}.  We can also exploit the predictive power of transfer entropy in identifying critical regimes of network dynamics and epidemic processes in particular \cite{sw11,e19050194}.  Precursors of phase transitions, identifiable by information dynamics \cite{ALife2011,PLOW11,spinney2017transfer} promise to strongly impact on the quality of epidemic forecasting.

Other future avenues of inquiry are aimed at developing a more versatile computational epidemiological framework by considering (i) more complicated outbreak scenarios, e.g., bio-terrorism attacks with {targeted infections}; (ii) different scales of modelling (fast and slow dynamics of disease diffusion); (iii) more diverse interaction types varying across demographic groups, professions, territories, seasons, and other contexts; and (v) compound intervention strategies, ranging from traditional vaccination to high-precision quarantine decisions.

\section{Acknowledgments}
The Authors are grateful to Tim Germann, Joseph Lizier, Cameron Zachreson, Kristopher Fair, David Newth, Philippa Pattison, Stephen Leeder, Tania Sorrell, Alessandro Montalto, and especially Peter Wang, for numerous helpful discussions of various intricacies involved in agent-based modelling of influenza. The Authors were supported through the Australian Research Council grant DP160102742.

\appendix

\section{Australian Census Data}

We are using data from the 2006 Australian Census obtained from the Australian Bureau of Statistics (ABS). All data contained in the this section can be obtained publicly, with the exception of the work travel data. It should be noted that some of the data needs to be processed using the \href{http://www.abs.gov.au/websitedbs/censushome.nsf/home/tablebuilder}{TableBuilder}. The census data has a hierarchical structure based on geography, described below.

\subsection{Hierarchical Structure of ABS data}

For the 2006 census, the ABS divided Australia using the Australian standard geographical classification~(ASGC). The levels present in the ASGC include: census collection district~(CD), local government area~(LGA), postal area~(POA), remoteness classification~(RA), statistical division~(SD), statistical local area~(SLA), state suburb~(SSC), statistical sub division~(SSD), urban centre or locality~(UCL), section of state~(SOS), state electoral division~(SED), commonwealth electoral division~(CED), state or territory~(STE). Some of these structures, such as RA and UCL, exist to express geographic concepts specific only to certain regions, such as remoteness and urbanity. Therefore these structures may not cover the entirety of Australia. In this work we use the following ASGC structures: CD, SLA, STE. In addition we also use the destination zone (DZN) classification. Each structure used in this work forms a complete partition of Australia. In addition, the census data in each CD is collected for each household, thus the household (HH) is also a \emph{de facto} structure of the census data.

\paragraph{State (STE)}
The STE is the highest level structure used by the ABS. This comprises the six states, two territories and one offshore territory. The offshore territory comprises any area which is not part of mainland Australia or Tasmania.

\paragraph{Statistical local area (SLA)}
The SLA is the second highest level structure used in this work, however this is only the fourth highest structure of the ASGC. In the 2006 census there are 1422 SLA's. The SLA has on average 14,064 people. The SLA's aggregate to states, that is no SLA is contained within two states.

\paragraph{Census collection district (CD)}
The CD is the structure with the highest hierarchical level used exclusively for the 'home regions' (see below). Data from a CD is collected by a single ABS representative, hence the name. In total there are 38,200 CDs which have an average population of 523 people. The CD's aggregate to SLA's, that is no CD is contained within two SLA's. 

\paragraph{Destination zone (DZN)}
The working group is the structure with the highest hierarchical level used exclusively for the 'work regions' (see below). The DZN describes the location that individuals go to work. In the 2006 census there are 9,800 DZN's. The DZNs aggregate to SLAs, i.e., no DZN is contained within two SLAs.

In addition, in our simulations we have defined and used some lower level structures which are relevant to the underlying infection dynamics, as described in the following section. These are the household cluster-HC and the working group~(WG).

\subsection{Australian Bureau of Statistics (ABS) data structures}

The ABS data is presented in tables. There are a number of tables which can be directly accessed from the ABS database. In addition, ABS provides the TableBuilder tool which we have used to build certain additional tables according to our simulation needs. It should be noted that the tables do not provide any data which could be used to identify individuals or households. Instead they are aggregated at the aforementioned structural levels (CD, SLA, STE). For example, a table built at the CD level could provide the number of individuals within the CD between certain age brackets, the number of households with a given number of members or the number of households who have a particular type of internet connection. The lowest structural level for which a table can be retrieved or built is the CD level. In this work, we retrieved or built the following tables for schools, family, households, age, sex, airport, travel to work.\footnote{The travel to work data is not publicly available and was purchased from the ABS.}

\paragraph{School distribution}
This was retrieved at the structural level of states (STE). This is the highest resolution at which the ABS offers data about schools. In this table, each row represents a state. Each column represents a range of the number of students attending the school. The states (rows) are NSW, VIC, QLD, SA, WA, TAS, NT, ACT. The ranges (columns) are 0-35, 36-100, 101-200, 201-300, 301-400, 401-600, 601-800, 801-1000, 1000+. 

\paragraph{Age distribution}
This was obtained at the structural level of the CD. In this table each row represents a CD. Each column gives the number of individuals of a specific age group and gender. The columns contain the number of males/females in the given CD who are between 0-4, 5-18, 19-39, 40-65, 65+.

\paragraph{Family distribution}
This table is constructed from three distinct tables from the ABS. The first table is defined at the level of CD and gives the number of households of a given size. The second table is also defined on the level of CD describes the number of each of four distinct types of family as defined by the ABS-Couples with children (CWC), couples without children (CWOC), single parent families (SPF) and other (OTH). The third table is defined on the national level gives the proportion of households which are family units or non family units. Non family units constitute households of either a single person, or people sharing accommodation without a familial relationship. Household Composition.dat is defined at the structural level of the CD. In this table each row represents a CD. Each column corresponds to a particular combination of household attributes derived from the first, second and third tables. Therefore each cell corresponds to the proportion of households with a particular combination of attributes in a given CD.

\paragraph{Worker flow}
Each row corresponds to a (CD, DZ) pair which represents a worker flow from a CD (source) to a DZ (target). The first column gives the CD code of the worker flow. The second column gives the DZ code of the flow and the third column gives the number of individuals living in the CD and working in the DZ.

\paragraph{Daily incoming passengers}
This table lists all the international airports in Australia. Each row in this table corresponds to a particular international airport and the number of passengers entering that airport from overseas per month on average. This data is taken from The Bureau of Infrastructure, Transport and Regional Economics (BITRE)~\cite{BITREairport}.

\section{Simulated population structure}

Following~\cite{GermannKadauEtAl2006,Longini2005} in constructing our national simulation model, we use a discrete-time stochastic simulation model of disease spread within a structured SLA. The model population is stochastically generated to match CD attributes for gender, age structure, household size, employment status, and location of employment among others.

\new{\acemod was implemented in C++11 and run on a High Performance Computing~(HPC) service (Artemis at the University of Sydney), comprising 4264 cores of computing capacity, consisting of 136 standard compute nodes, two high memory nodes (512 GB), three very high memory nodes (6 TB), and five GPU compute nodes, under a ``Fair Share'' resource allocation model (e.g., no more than 288 cores). A typical run of 180 days of an epidemic is simulated on an HPC node in 42 minutes, so that each scenario covering 10 runs varying over infection ``seeds'' is complete and integrated within seven hours. However, the code is not optimised for performance and includes numerous debugging statistics; thus much lower run-times could be achieved with minimal effort.}

Each person in the population belongs to one of five age groups: preschool-age children (0--4 years), school-age children (5--18 years), young adults (19--29 years), adults (30--64 years), and older adults (64+ years). The individuals within the model are arranged into one of 4 family structures: singles; couples without children; couples with children; and single parent families.   In general, households consist of a number of persons, matching the Australian census data, and are grouped randomly into clusters of four households each.

Every person also belongs to a set of close and casual contact (also referred to as ``mixing'') groups, ranging from their household and household cluster (highest contact probabilities) to workplaces, collection districts, and to the SLA (with the lowest contact probabilities, representing occasional ``long distance'' interactions in supermarkets, shopping centres, theatres, libraries, mass gatherings, etc.). 

Depending on their age, school-age children attend primary school or high-school.  In each case, children are arranged into 25 person groups referred to as `classes'. According to the Australian Census data 98.7\% of children 5-18 years old attend school, so we allow the remaining 1.3\% to mix in the household, household cluster, and community during daytime. Working adults (restricted to those who are 19--64 years old) belong to a working group of a varying size determined by census data. We assume that workers make a contact of sufficient duration and/or closeness to transmit the disease with a subset of the entire workforce at that location. 

\subsection{Assigning nighttime mixing groups} \label{app:home}

In order to populate the nighttime (home) regions, our simulation iterates through CDs from the census data. For each CD, a cumulative density function~(CDF) is constructed from the family distributions as described above. For a particular CD, we then create households one by one. Each household has attributes assigned to it probabilistically based on the CDF, including the number of people living in the household and the type of family living in that household. This process continues until the population of the CD is reached. After the household attributes are chosen, individuals are assigned their own attributes, e.g., a single parent family of size 3 requires one adult and two children. Therefore one person will be assigned as an adult, and two will be assigned as children. The age and gender of each individual is then randomly assigned using data from the age distribution.

\subsection{Assigning daytime mixing groups} \label{app:work}

After home regions have been populated, the worker flow information is used to generate mobility data. This data gives the number of workers travelling between a CD and a DZN. This number of workers are then selected randomly from the CD and assigned as workers in the given DZN. Therefore for each (CD, DZN) pair, we randomly select the given number of unassigned, working aged individuals from the corresponding CD and assign them to the corresponding DZN.  Figure~\ref{fig:commute-dist} shows the distribution of commute distances for working individuals, as well as school students (described in the next subsection), generated by \acemod. 

\begin{figure}
	\centering
	\subfigure[Workers]{ {\includegraphics[width=0.45\textwidth]{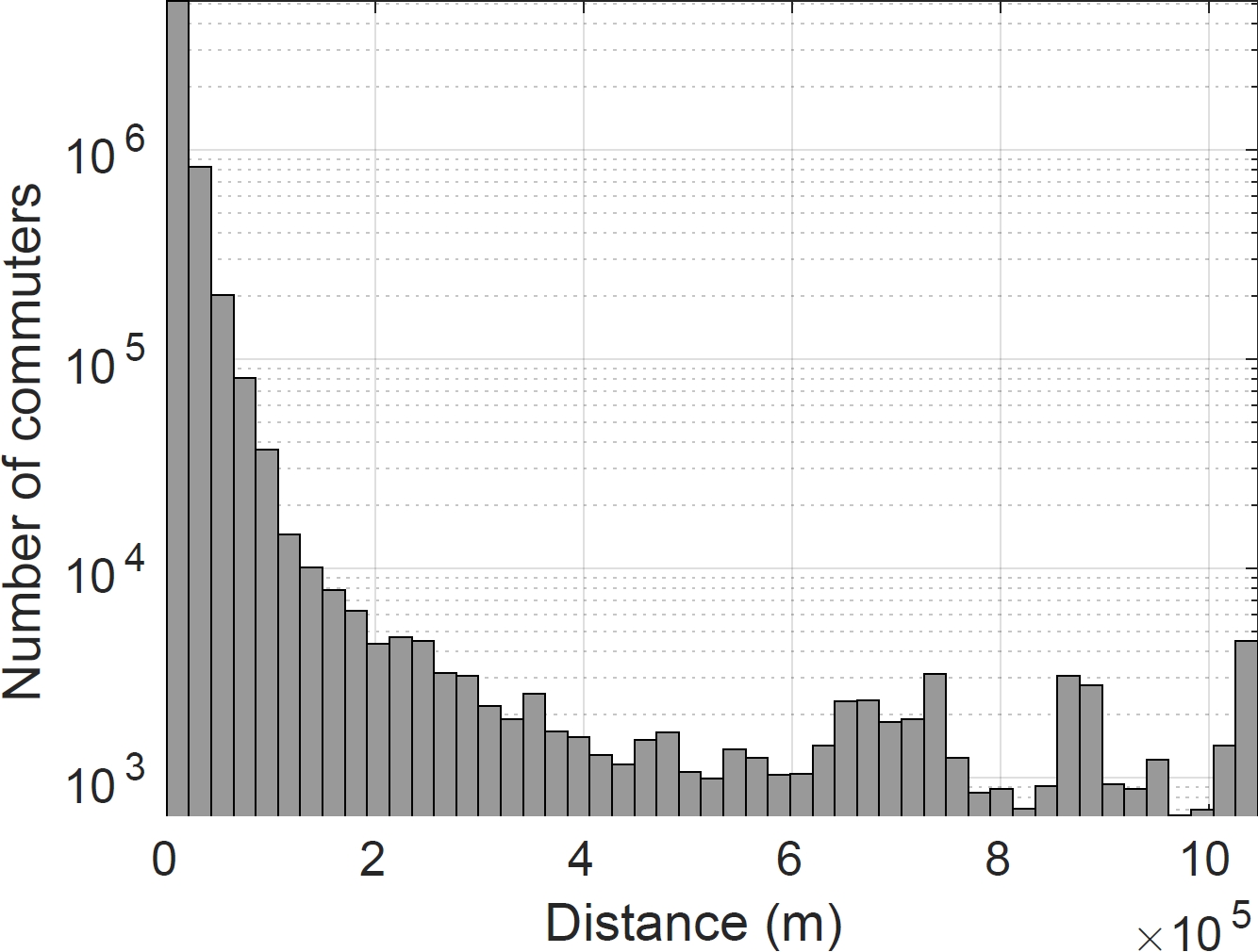}} \label{fig:worker-commute} }\hfill
	\subfigure[Students]{ {\includegraphics[width=0.45\textwidth]{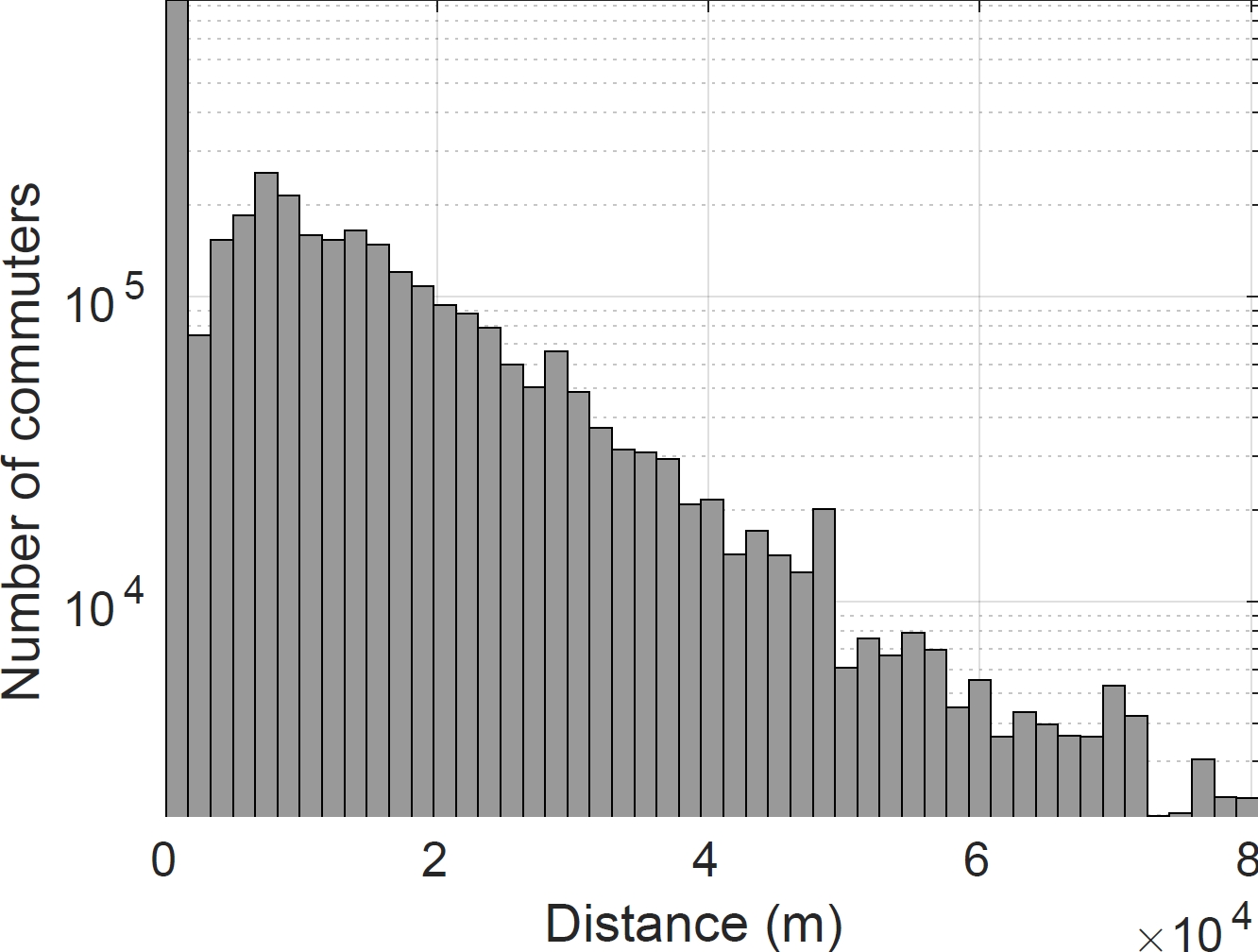}} \label{fig:student-commute} } 
	\caption{Commute distance distributions.}\label{fig:commute-dist}
\end{figure}

\subsubsection{Populating schools} \label{app:schools}

Our simulation generates schools and assigns a maximum capacity to these schools according to the distribution of school sizes available from ABS. The schools are then spatially assigned based on the availability of school staff in the relevant DZNs. The number of school staff in each DZN will determine the number of students able to go to school in that DZN based on a fixed staff to student ratio. Each school is randomly assigned to a DZN which has a sufficient number of  potential students. The number of potential students in a DZN is then reduced by the size of the school. This continues until all potential schools have been allocated to a DZ. At this point, each DZN has a number of schools with unfilled student and staff places.

In order to assign students to schools, we assume that students would go to a school if they live within the catchment zone of the school.  This process mimics the process that students are likely to attend a schools within a short distance first, and if there are no places available they would search further afield. Similarly, schools are more likely to accept students from a greater distance if it is evident that there are no other closer options. If no school could be found within the maximum radius, it is assumed that the student does not attend a school. The catchment area of a school is defined by randomly selecting a radius from a distribution, which is a folded normal distribution converted from a normal distribution with mean and variance of 10km and 4km. If the student lives within the catchment zone of many schools, the algorithm will pick one school with probability proportional to the remaining available student places in the school. Once a student place is chosen, the total available student places in the given school are reduced by one. We do not distinguish between public and private schools for the allocation purposes of students. If a student does not live within a catchment zone of any school, the algorithm searches an area with double the original radius and again selects a school with probability proportional to the available places. If there is still no school within the doubled radius, the radius is doubled again. This process is repeated until the algorithm finds a school or reaches the maximum searching radius of 100 km.

Once the students are all assigned to school places, we use the student to staff ratio (2:17) (extracted from historical ABS data) and the enrolment numbers of the schools to calculate the number of staff working in each school. These staff members are then selected randomly from the ensemble of people who work in that particular DZN. The remaining people who work in the DZN are assigned to non-school working groups, as follows. The number of remaining workers is divided by the maximum size of a working group (20) and the ceiling of this number is the number of non school working groups within the particular DZN. Workers are then randomly assigned to these working groups such that none will exceed a size of 20.

The results of this assignment algorithm are summarised in the following figures: 
\begin{itemize}
	\item Figure~\ref{fig:school-sizes} shows school size distributions;
	\item Figure~\ref{fig:commute-dist} show commute to school (gamma) distribution (mean distance to school).
\end{itemize}

\begin{figure}
	\centering
	\subfigure[NSW]{{\includegraphics[width=.325\textwidth]{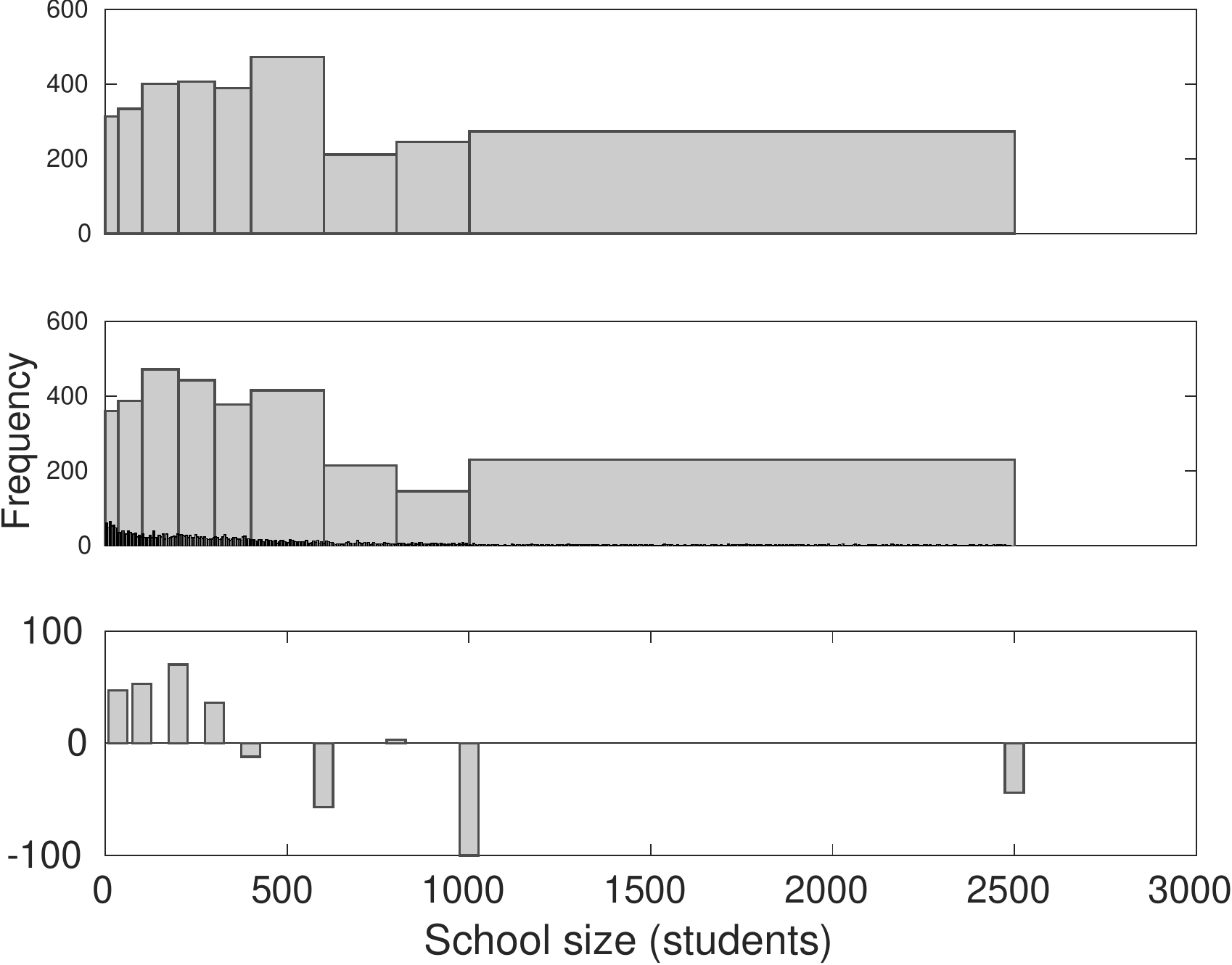}}\label{fig:school-dist-NSW}} \hspace{10mm}
	\subfigure[VIC]{{\includegraphics[width=.325\textwidth]{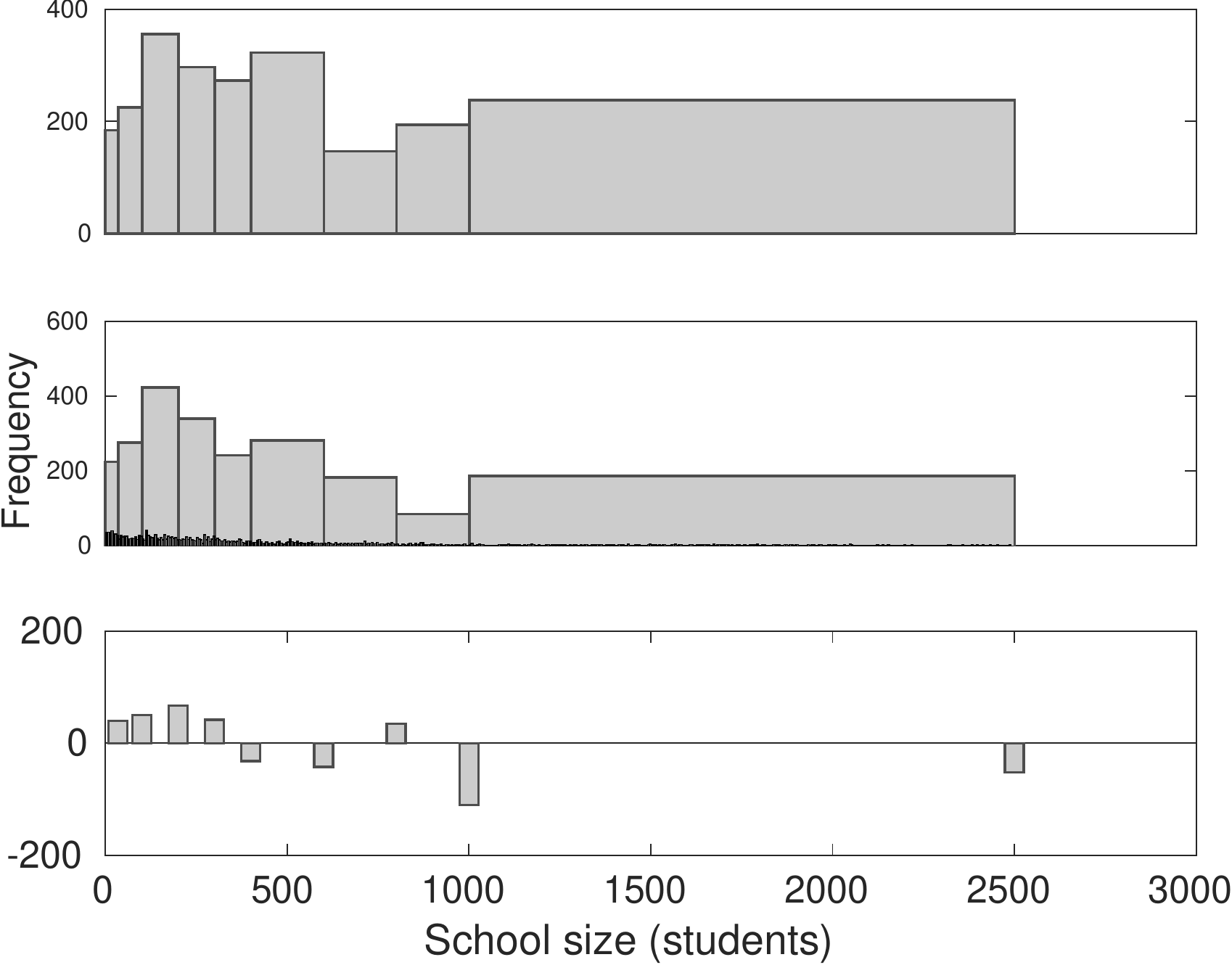}}\label{fig:school-dist-VIC}} \\
	\subfigure[QLD]{{\includegraphics[width=.325\textwidth]{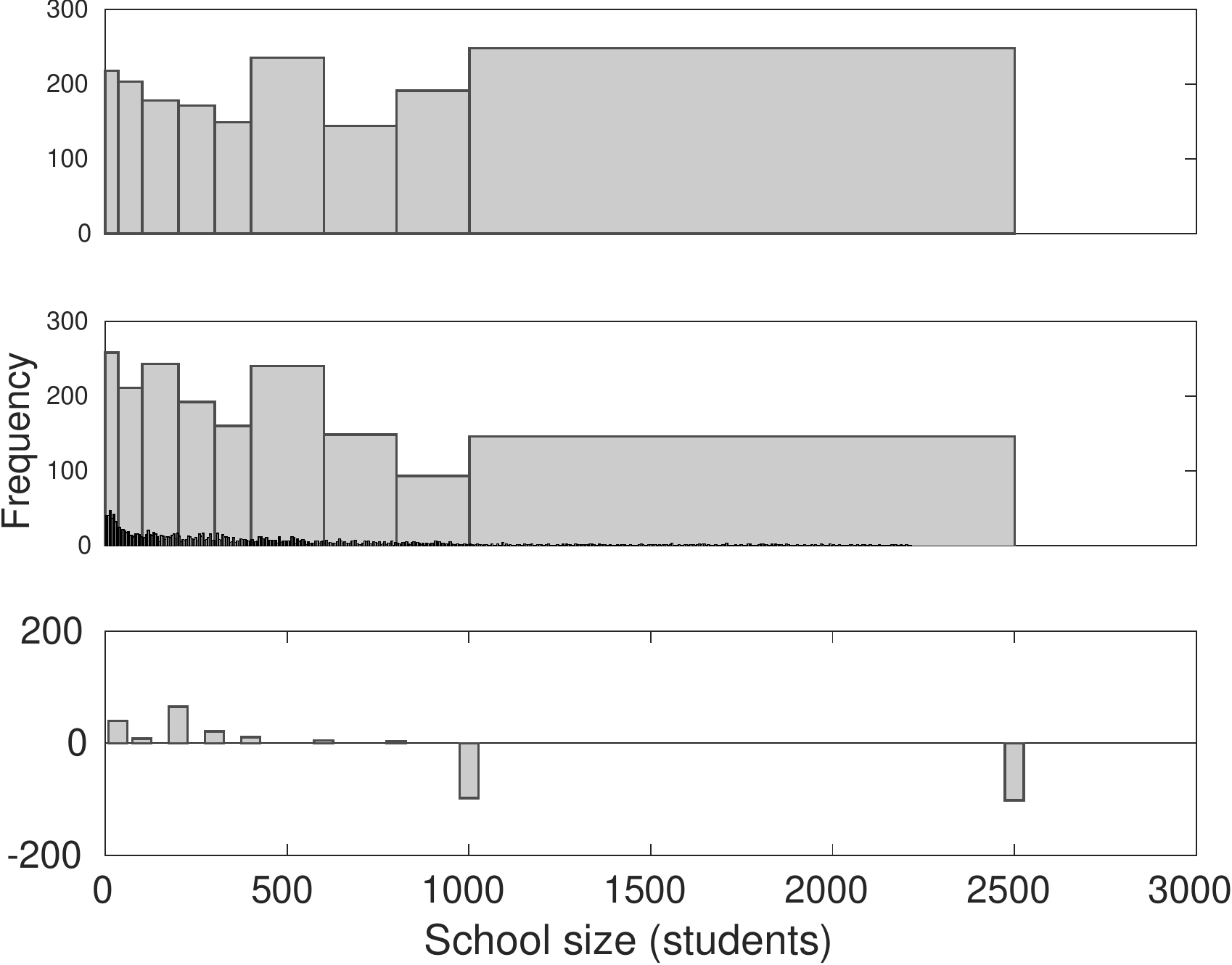}}\label{fig:school-dist-QLD}} \hspace{10mm}
	\subfigure[WA]{{\includegraphics[width=.325\textwidth]{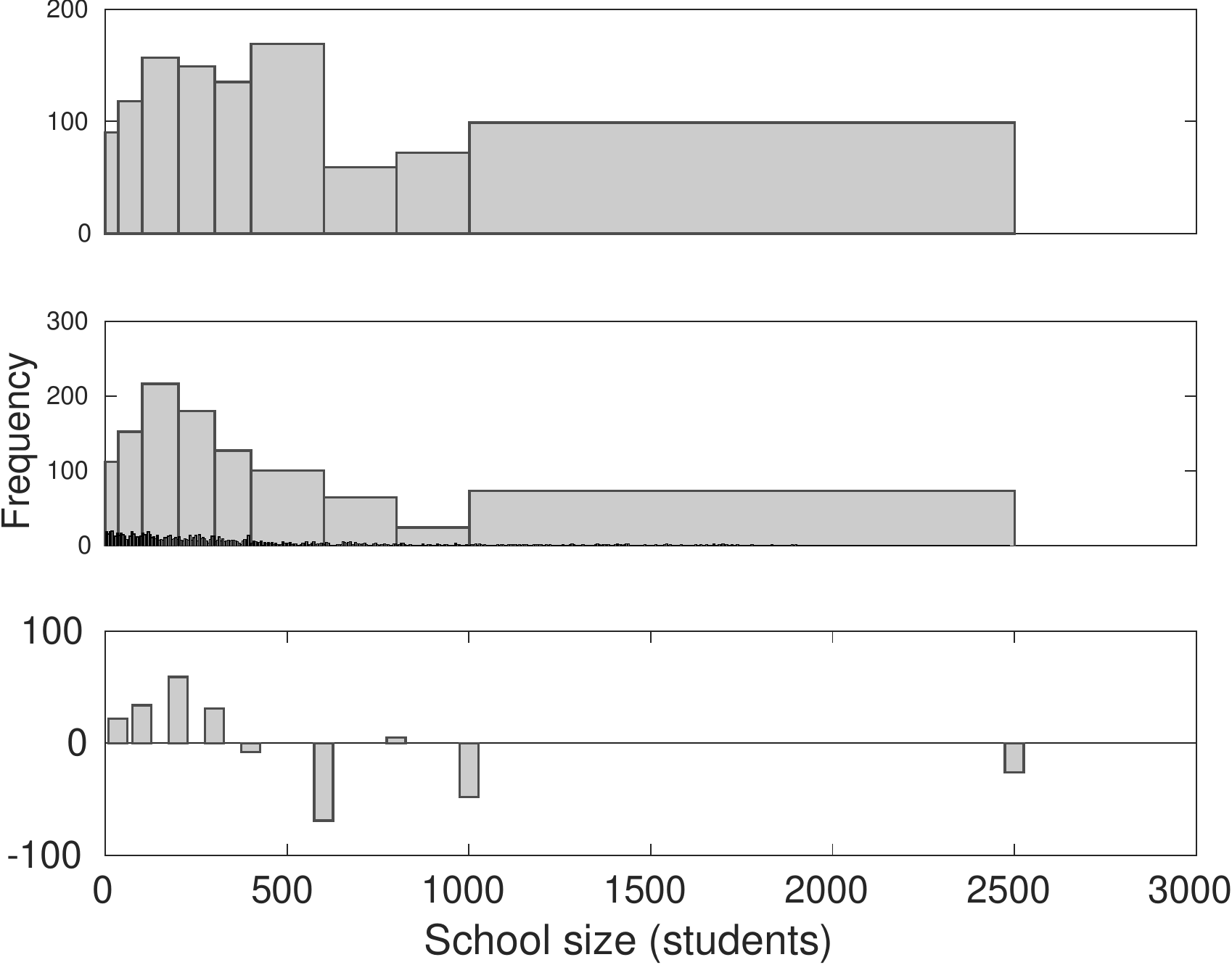}}\label{fig:school-dist-WA}} \\
	\subfigure[NT]{{\includegraphics[width=.325\textwidth]{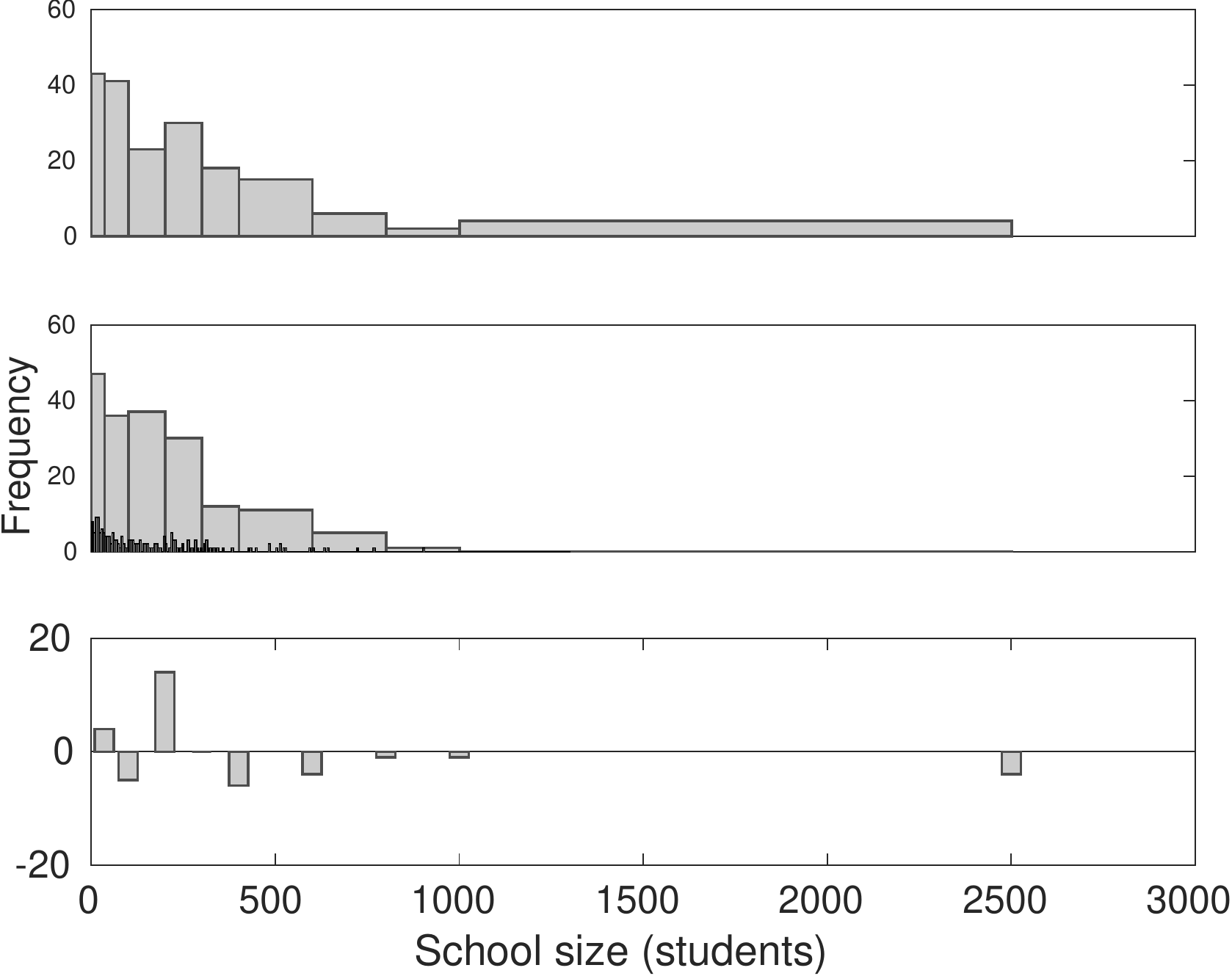}}\label{fig:school-dist-NT}} \hspace{10mm}
	\subfigure[SA]{{\includegraphics[width=.325\textwidth]{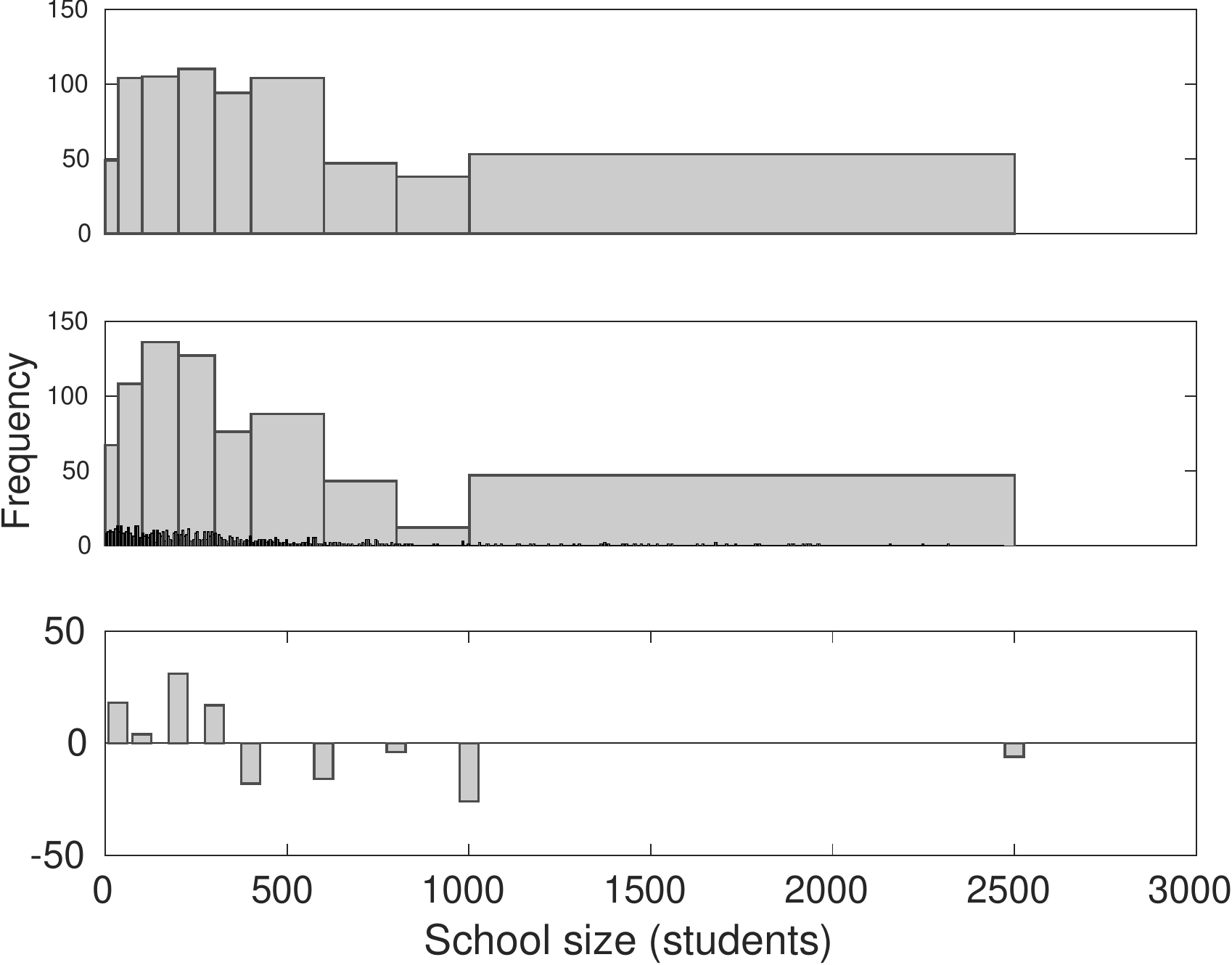}}\label{fig:school-dist-ACT}} \\
	\subfigure[TAS]{{\includegraphics[width=.325\textwidth]{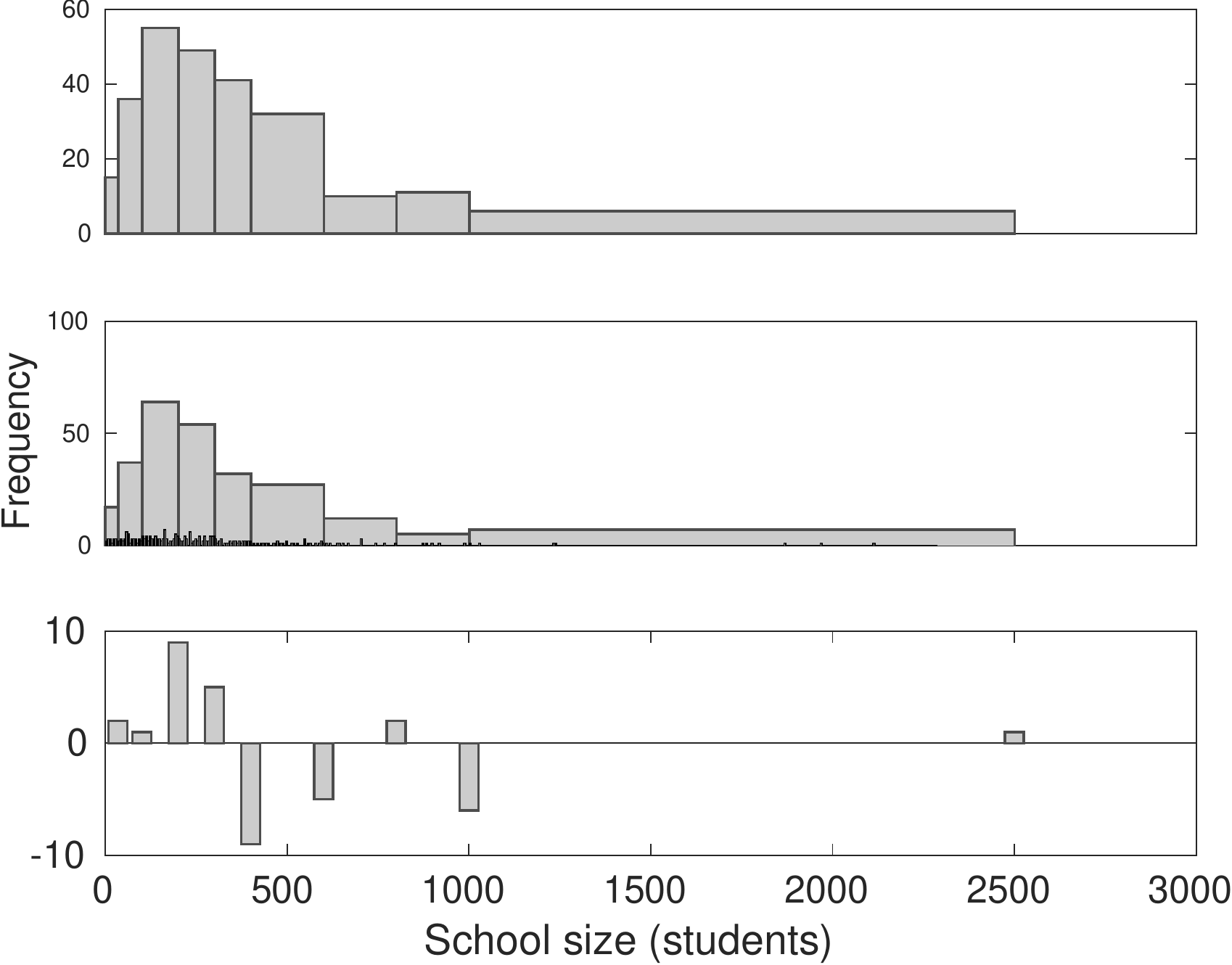}}\label{fig:school-dist-TAS}} \hspace{10mm}
	\subfigure[ACT]{{\includegraphics[width=.325\textwidth]{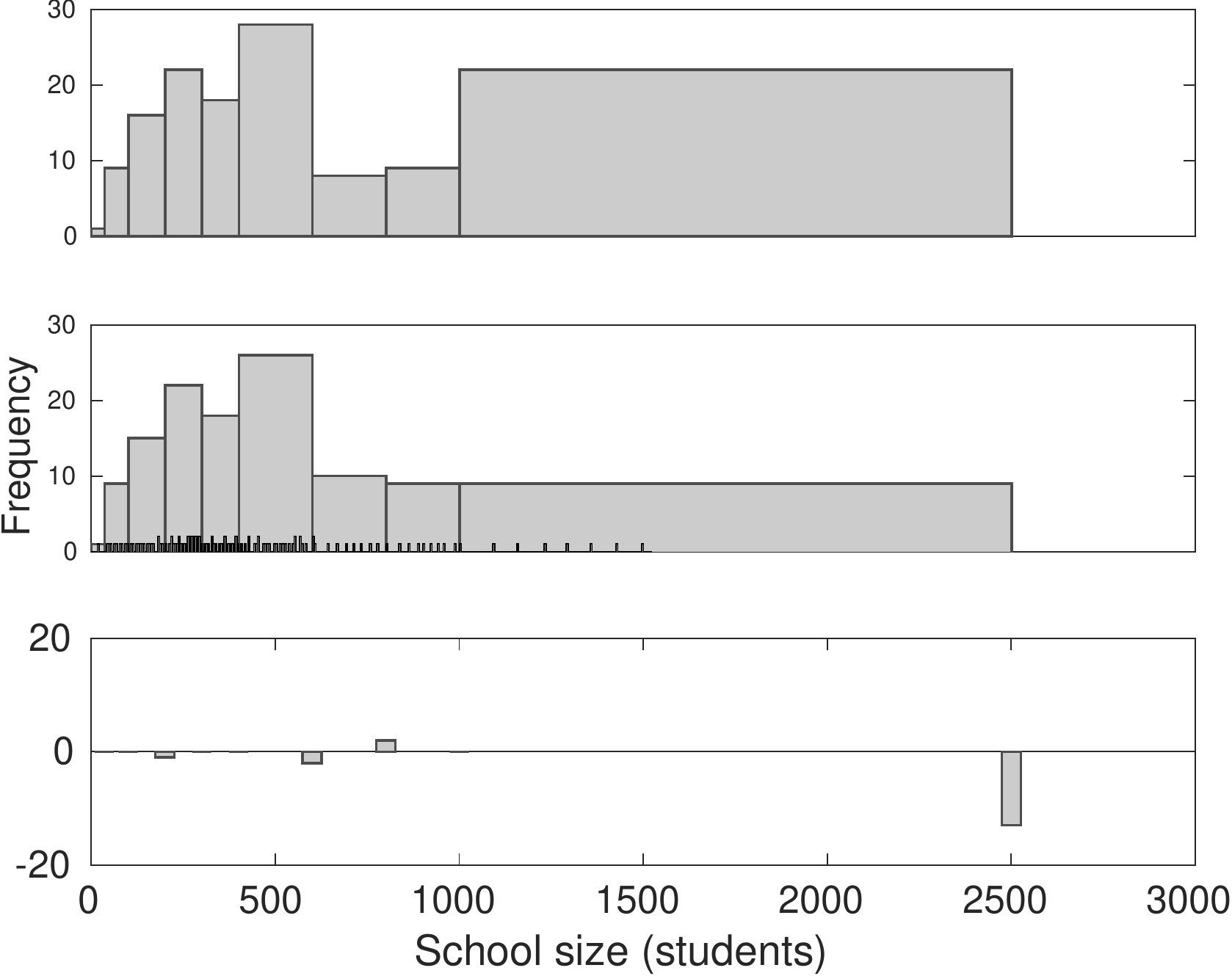}}\label{fig:school-dist-ACT}}
	\caption{School size distributions.}\label{fig:school-sizes}
\end{figure}

\section{Transmission and contact probabilities}

The transmission probabilities (for Eq.~\eqref{eq:cauch}) and contact probabilities (for Eq.~\eqref{eq:prob-transmission}) are given in Tab.~\ref{tab:contact-table} and Tab.~\ref{tab:transmission-rates}, respectively.

\bgroup
\begin{table}
	\caption{Daily contact probabilities $c_{j \to i}^g$ for different contact groups $g$, reported by~\cite{chao2010flute}.}
	\label{tab:contact-table}
	\vspace{1mm}
	\centering
	\rowcolors{1}{white}{gray!5!}
\resizebox{.8\textwidth}{!}{%
	{\raggedright
	 \noindent
	\begin{tabular}{llll}
	Mixing group $g$ & Infected individual $j$ & Susceptible individual $i$ & Contact probability $c_{j \to i}^g$ \\
	\hline
	Household cluster & Child (\textless19) & Child (\textless19) &  0.08 \\
	 & Child (\textless19) & Adult (\textgreater18) & 0.035 \\
	 & Adult (\textgreater18) & Child (\textless19) & 0.025  \\
	 & Adult (\textgreater18) & Adult & 0.04 \\
	\hline
	Working Group & Adult (19-64) & Adult (19-64) & 0.05 \\
	\hline
	Neighbourhood & Any & Child (0-4) &  0.0000435 \\
	 & Any & Child (5-18) & 0.0001305 \\
	 & Any & Adult (19-64) & 0.000348  \\
	 & Any & Adult (65+) & 0.000696 \\
	\hline
	Community & Any & Child (0-4) &  0.0000109 \\
	 & Any & Child (5-18) & 0.0000326 \\
	 & Any & Adult (19-64) & 0.000087  \\
	 & Any & Adult (65+) & 0.000174 \\
	\hline
	\end{tabular}
	}
}
\end{table}
\egroup

\bgroup
\def\arraystretch{1.3}
\setlength\arrayrulewidth{1pt}
\setlength\tabcolsep{4mm}
\begin{table}
\centering
\rowcolors{1}{white}{gray!5!}
 \noindent
 \caption{Daily transmission {probabilities} $q_{j\to i}^g$ for different contact groups $g$, obtained by Eq.~\eqref{eq:cauch} where $\beta_{j\to i}^g$ are reported by~\cite{cauchemez2011role}.
 }
 \label{tab:transmission-rates}
 \vspace{1mm}
 \resizebox{.8\textwidth}{!}{%
 	{\raggedright
\begin{tabular}{llll}
Contact Group $g$ & Infected Individual $j$ & Susceptible Individual $i$ & Transmission Probability $q^g_{j \to i}$ \\
\hline

Household size 2 & Any & Child (\textless19) & 0.0933 \\
& Any & Adult (\textgreater18) & 0.0393 \\
\hline
Household size 3 & Any &  Child (\textless19) & 0.0586 \\
 & Any & Adult (\textgreater18) & 0.0244 \\
\hline
Household size 4 & Any & Child (\textless19) & 0.0417 \\
& Any & Adult (\textgreater18) & 0.0173 \\
\hline
Household size 5 & Any & Child (\textless19) & 0.0321 \\
 & Any & Adult (\textgreater18) & 0.0133 \\
\hline
Household size 6 & Any & Child (\textless19) & 0.0259 \\
& Any &  Adult (\textgreater18) & 0.0107 \\
\hline
School & Child (\textless19) & Child (\textless19) & 0.000292 \\
Grade & Child (\textless19) & Child (\textless19) & 0.00158 \\
Class & Child (\textless19) & Child (\textless19) & 0.035 \\
\hline
\end{tabular}
}}
\label{transmission_table}
\end{table}
\egroup

\section{Supplementary results}

Figure~\ref{fig:r0-hists} illustrates the number of secondary cases from a typical outbreak. Figure~\ref{fig:spatial-syd} shows the prevalence choropleth for the Greater Sydney Area.

\begin{figure}
	\centering
	\subfigure[$\kappa = 1.0$]{{\includegraphics[width=.35\textwidth]{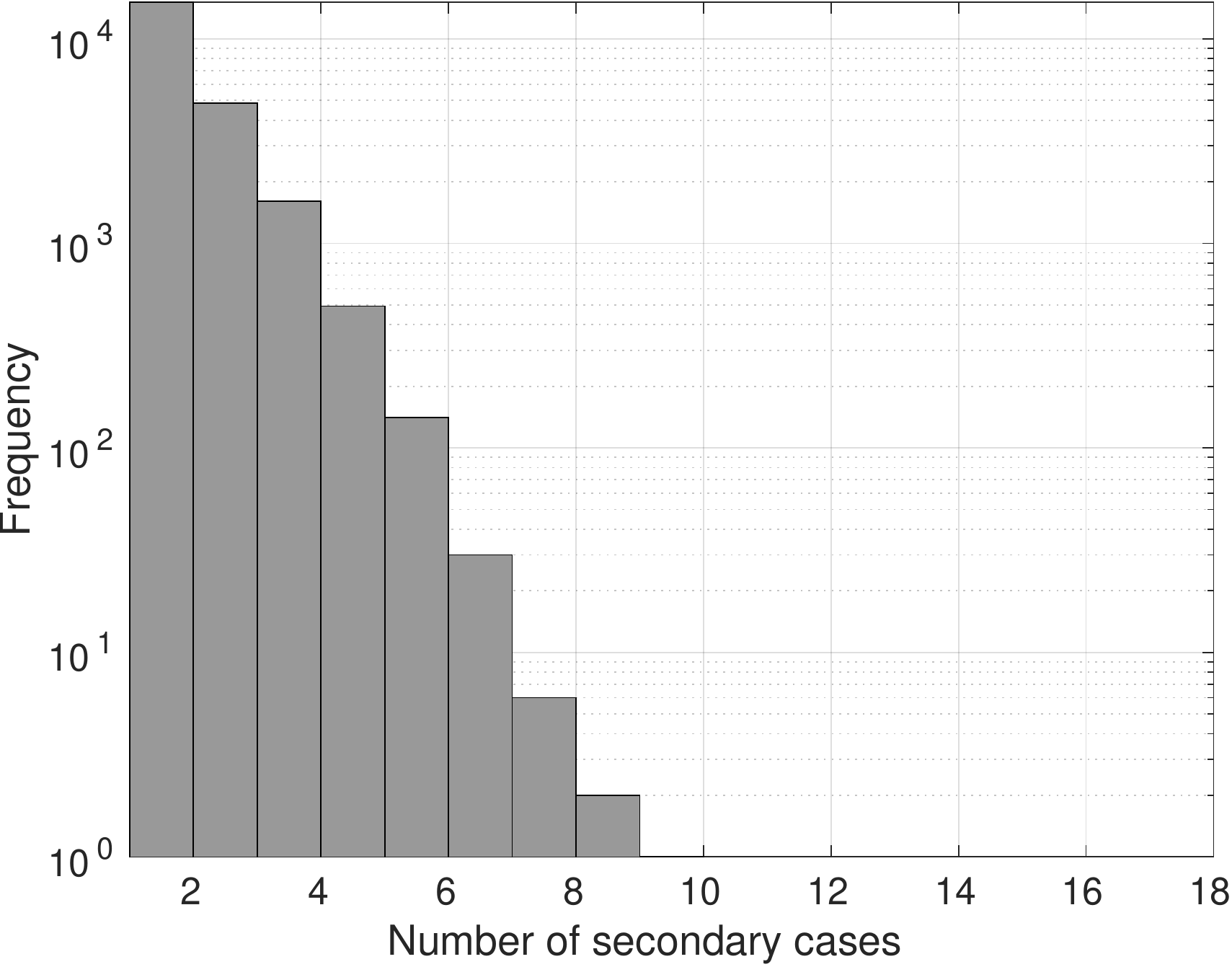}}\label{fig:r0_1}} \hspace{10mm}
	\subfigure[$\kappa = 2.0$]{{\includegraphics[width=.35\textwidth]{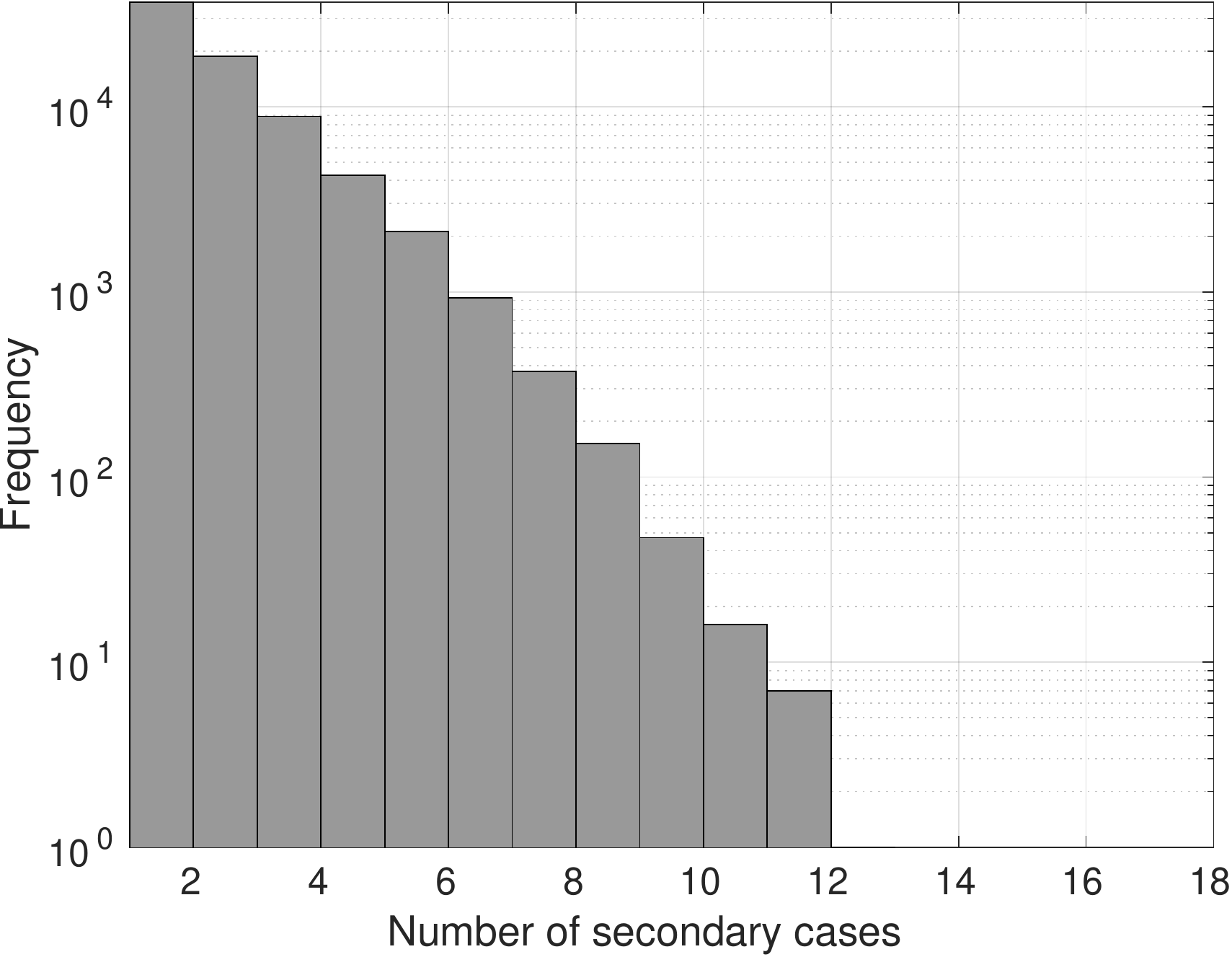}}\label{fig:r0_2}} \\
	\subfigure[$\kappa = 3.0$]{{\includegraphics[width=.35\textwidth]{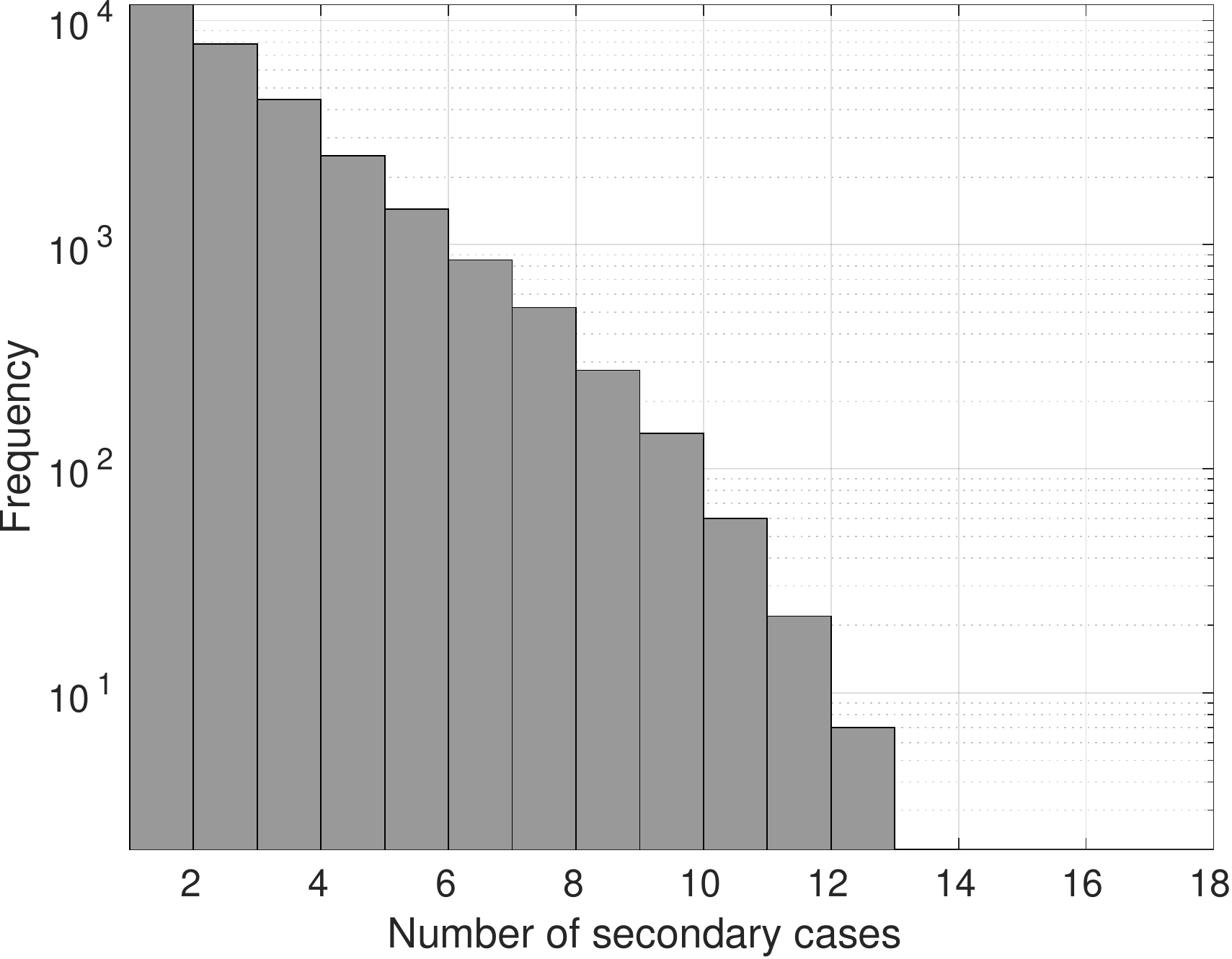}}\label{fig:r0_3}} \hspace{10mm}
	\subfigure[$\kappa = 4.0$]{{\includegraphics[width=.35\textwidth]{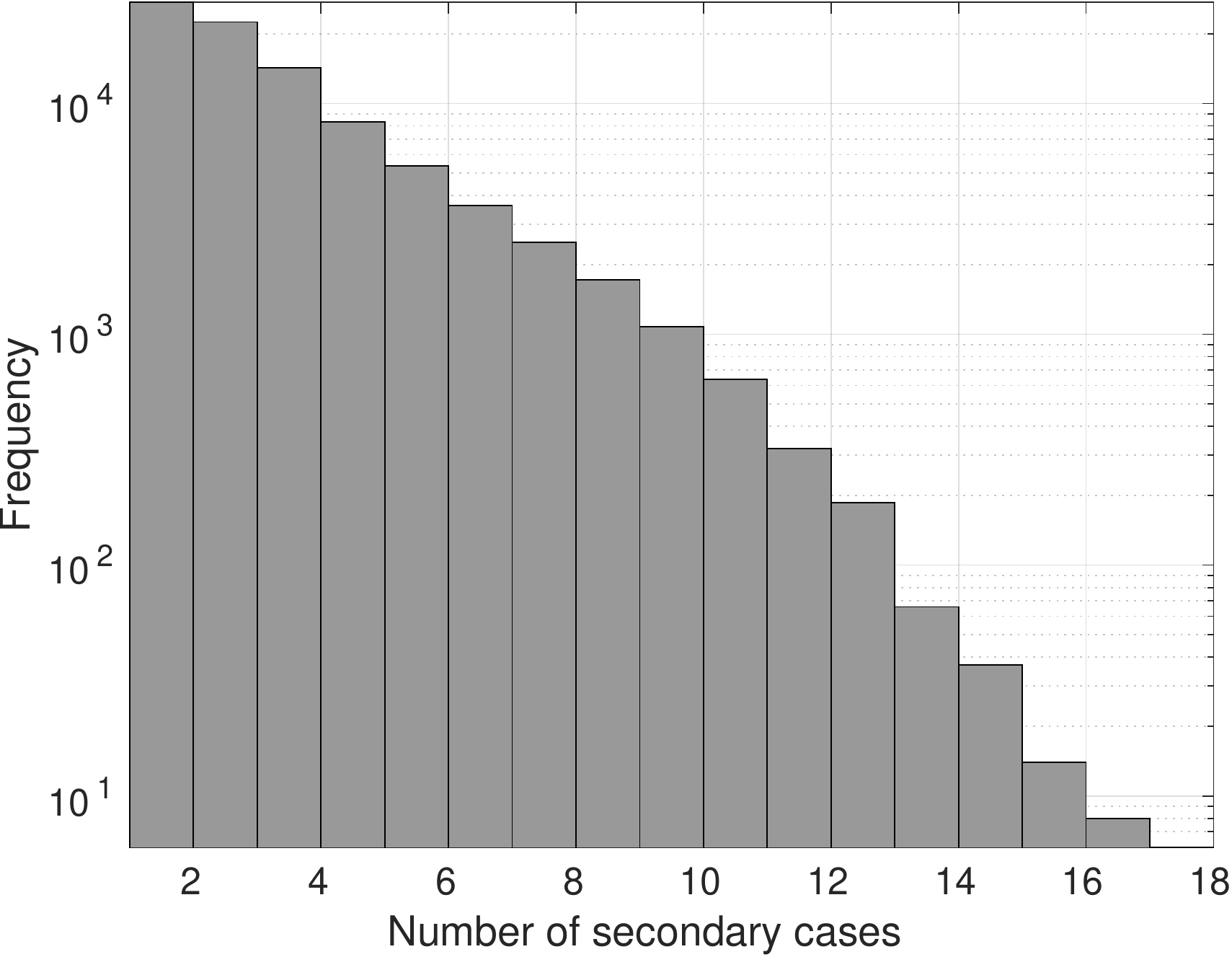}}\label{fig:r0_4}}
	\caption{Number of secondary cases for simulations with $\kappa = \{1.0, 2.0, 3.0, 4.0\}$.}\label{fig:r0-hists}
\end{figure}


\begin{figure}
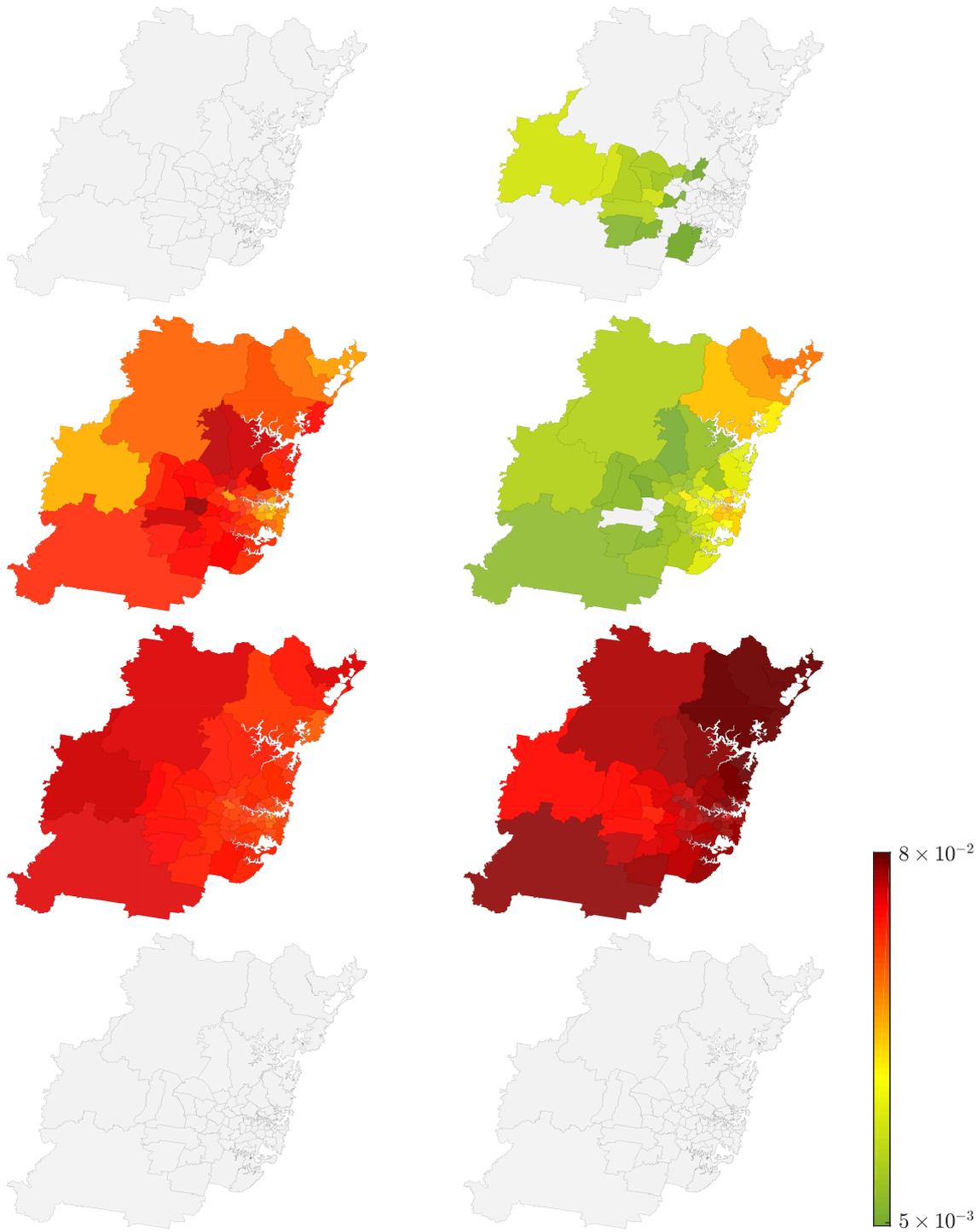

	\centering
	\subfigure{{\includegraphics[height=190mm]{P_SLA_R1-5_SYD}}\label{fig:p_sla_r1-75}} \hspace{12.5mm}
  \subfigure{{\includegraphics[height=190mm]{P_SLA_R2_SYD}}\label{fig:p_sla_r2}} \hspace{5mm}
  \subfigure{{\includegraphics[height=60mm]{heatmap_colorbar-crop}}\label{fig:cbar}} \hfill\null
	\caption{Prevalence proportion choropleths of Sydney for $R_0 = 1.5$ and $R_0 = 2$. We plot the distribution for days 30, 50, 62, and 88, with time going down the page. Both simulations are realisations comprising the same demographics (contact) and mobility networks, as well as identical seeding	at the same rate at major international airports around Australia (see Tab.~\ref{tab:airport}).} \label{fig:spatial-syd}
\end{figure}

\clearpage


\bibliographystyle{plain}

\end{document}